\documentclass[twocolumn,fleqn]{svjour2}    
\usepackage{graphicx}
\usepackage{graphicx}
\usepackage{dcolumn}
\usepackage{bm}
\usepackage{amsmath}
\usepackage{graphicx}

\def\gsim{\;\lower4pt\hbox{${\buildrel\displaystyle >\over\sim}$}\;}
\def\lsim{\;\lower4pt\hbox{${\buildrel\displaystyle <\over\sim}$}\;}
\def\grls{\;\lower4pt\hbox{${\buildrel\displaystyle >\over <}$}\;}

\journalname{Astrophysics and Space Science}
\begin{document}

\title{Dynamic Evolution Model of Isothermal Voids and Shocks}

\author{Yu-Qing Lou  \and Xiang Zhai }


\institute{Y.-Q. Lou \at
           1. Physics Department and Tsinghua Center for
Astrophysics (THCA), Tsinghua University, Beijing 100084, China;\\
2. Department of Astronomy and Astrophysics, The University
of Chicago, 5640 S. Ellis Avenue, Chicago, IL 60637 USA;\\
3. National Astronomical Observatories, Chinese Academy
of Sciences, A20, Datun Road, Beijing 100012, China.\\
\email{louyq@tsinghua.edu.cn; lou@oddjob.uchicago.edu}
        \and X. Zhai \at
              Physics Department and Tsinghua Center
for Astrophysics (THCA), Tsinghua University, Beijing 100084,
China\\  \email{zxzhaixiang@gmail.com} }

\date{Received: date / Accepted: date}

\maketitle

\begin{abstract}
We explore self-similar hydrodynamic evolution of central voids
embedded in an isothermal gas of spherical symmetry under the
self-gravity. More specifically, we study voids expanding at
constant radial speeds in an isothermal gas and construct all types
of possible void solutions without or with shocks in surrounding
envelopes. We examine properties of void boundaries and outer
envelopes. Voids without shocks are all bounded by overdense shells
and either inflows or outflows in the outer envelope may occur.
These solutions, referred to as type $\mathcal{X}$ void solutions,
are further divided into subtypes $\mathcal{X}_{\rm I}$ and
$\mathcal{X}_{\rm II}$ according to their characteristic behaviours
across the sonic critical line (SCL). Void solutions with shocks in
envelopes are referred to as type $\mathcal{Z}$ voids and can have
both dense and quasi-smooth edges. Asymptotically, outflows,
breezes, inflows, accretions and static outer envelopes may all
surround such type $\mathcal{Z}$ voids. Both cases of constant and
varying temperatures across isothermal shock fronts are analyzed;
they are referred to as types $\mathcal{Z}_{\rm I}$ and
$\mathcal{Z}_{\rm II}$ void shock solutions. We apply the `phase net
matching procedure' to construct various self-similar void
solutions. We also present analysis on void generation mechanisms
and describe several astrophysical applications. By including
self-gravity, gas pressure and shocks, our isothermal self-similar
void (ISSV) model is adaptable to various astrophysical systems such
as planetary nebulae, hot bubbles and superbubbles in the
interstellar medium as well as supernova remnants. \keywords{H II
regions \and hydrodynamics \and ISM: bubbles \and ISM: clouds \and
planetary nebulae \and supernova remnants}
\PACS{95.30.Qd\and98.38.Ly\and95.10.Bt\and97.10.Me\and97.60.Bw}
\end{abstract}

\section[]{Introduction}\label{intro}
\renewcommand{\topfraction}{0.98}

Voids may exist in various astrophysical cloud or nebula systems of
completely different scales, such as planetary nebulae,
supernova remnants, interstellar bubbles and superbubbles and so
forth.
In order to understand the large-scale dynamic evolution of such
gaseous systems, we formulate and develop in this paper an
isothermal self-similar dynamic model with spherical symmetry
involving central expanding voids under the self-gravity.
In our dynamic model framework, a void is approximately regarded as
a massless region with a negligible gravity on the surrounding gas
envelope. With such idealization and simplification in our
hydrodynamic equations, a void is defined as a spherical space or
volume containing nothing inside. In any realistic astrophysical
systems, there are always materials inside voids such as stellar
objects and stellar winds or outflows etc. In Section 4.2.3, we
shall show that in astrophysical gas flow systems that our
isothermal self-similar void (ISSV) solutions are generally
applicable, such as planetary nebulae, interstellar bubbles or
superbubbles and so on.


Observationally, early-type stars are reported to blow strong
stellar winds towards the surrounding interstellar medium (ISM).
Hydrodynamic studies on the interaction of stellar winds
with surrounding gases have shown that a stellar wind will sweep up
a dense circumsteller shell (e.g. Pikel'ner \& Shcheglov 1968;
Avedisova 1972; Dyson 1975; Falle 1975). Such swept-up density
`wall' surrounding a central star thus form interstellar bubbles of
considerably low densities inside (e.g. Castor, McCray \& Weaver
1975).
For example, the Rosette Nebula (NGC 2237, 2238, 2239, 2246)
is a vast cloud of dusts and gases spanning a size of $\sim 100$
light years.
It has a spectacular appearance with a thick
spherical shell of ionized gas and a cluster of luminous massive OB
stars in the central region,
whose strong stellar winds and intense radiation have
cleared a `hole' or `cavity'
around the centre and given rise to a thick spherical shell of
ionized gases (e.g. Mathews 1966; Dorland, Montmerle \& Doom 1986).
Weaver et al. (1977) outlined a dynamic theory to explain
interstellar bubbles. They utilized equations of motion and
continuity with spherical symmetry. They gave an adiabatic
similarity solution, which is applicable at early times and also
derived a similarity solution including the effect of thermal
conduction between the hotter (e.g. $T\approx 10^6$K) interior and
the colder shell of swept-up ISM. Their solution was also modified
to include effects of radiation losses. Weaver et al. (1977) did not
consider the self-gravity of gas shell which can be dynamically
important for such a large nebula,
and therefore possible
behaviours of their self-similar solutions were fairly limited. For
example, the thickness of the gas shell was limited to $\sim 0.14$
times of the bubble radius. In our model formulation, we ignore the
gravity of the central stellar wind region and of star(s) embedded
therein. Thus, this central region is treated as a void and we
explore the self-similar dynamic behaviours of surrounding gas shell
and ISM involving both the self-gravity and thermal pressure. Our
ISSV solutions reveal that the gas shell of a cloud can have many
more types of dynamic behaviours.

%

Planetary nebulae (PNe) represent an open realm of astrophysical
applications of our ISSV model, especially for those that appear
grossly spherical (e.g. Abell 39; see also Abell
1966).\footnote{Most planetary nebulae appear elliptical or bipolar
in terms of the overall morphology.}
During the stellar
evolution, planetary nebulae emerge during the transition from the
asymptotic giant branch (AGB) phase where the star has a slow AGB
dense wind to a central compact star (i.e. a hot white dwarf) where
it blows a fast wind in the late stage of stellar evolution (e.g.
Kwok, Purton \& Fitzgerald 1978; Kwok \& Volk 1985; Chevalier
1997a). The high temperature of the compact star can be a source of
photoionizing radiation and may partially or completely photoionize
the dense slower wind. Chevalier (1997a) presented an isothermal
dynamical model for PNe and constructed spherically symmetric global
hydrodynamic solutions to describe the expansion of outer shocked
shell with an inner contact discontinuity of wind moving at a
constant speed.
In Chevalier (1997a), gravity is ignored
and the gas flow in the outer region can be either winds or breezes.
In this paper, we regard the inner expansion region of fast wind as
an effective void and use ISSV solutions with shocks to describe
dense shocked wind and AGB wind expansion. One essential difference
between our ISSV model and that of Chevalier (1997a) lays in the
dynamic behaviour of the ISM. In Chevalier (1997a), shocked envelope
keeps expanding with a vanishing terminal velocity or a finite
terminal velocity at large radii. By including the self-gravity, our
model can describe a planetary nebula expansion surrounded by an
outgoing shock which further interacts with a static, outgoing or
even accreting ISM. In short, the gas self-gravity is important to
regulate dynamic behaviours of a vast gas cloud. Quantitative
calculations also show that the lack of gas self-gravity may lead to
a considerable difference in the void behaviours (see Section 4.1).
Likewise, our ISSV model provides more sensible results than those
of Weaver et al. (1977). We also carefully examine the inner fast
wind region and show that a inner reverse shock must exist and the
shocked fast wind has a significant lower expansion velocity than
the unshocked innermost fast wind. It is the shocked wind that
sweeps up the AGB slow wind, not the innermost fast wind itself.
This effect is not considered in Chevalier (1997a). We also compare
ISSV model with Hubble observations on planetary nebula NGC 7662 and
show that our ISSV solutions are capable of fitting gross features
of PNe.

Various aspects of self-similar gas dynamics have been investigated
theoretically for a long time (e.g. Sedov 1959; Larson 1969a, 1969b;
Penston 1969a, 1969b; Shu 1977; Hunter 1977, 1986; Landau \&
Lifshitz 1987; Tsai \& Hsu 1995; Chevalier 1997a; Shu et al 2002;
Lou \& Shen 2004; Bian \& Lou 2005). Observations also show that gas
motions of this kind of patterns may be generic.
Lou \& Cao
(2008) illustrated one general polytropic example of central void in
a self-similar expansion as they explored self-similar dynamics of a
relativistically hot gas with a polytropic index $4/3$ (Goldreich \&
Weber 1980; Fillmore \& Goldreich 1984).
The conventional polytropic gas model of Hu \& Lou (2008)
considered expanding central voids embedded in ``champagne flows" of
star forming clouds and provided versatile solutions to describe
dynamic behaviours of ``champagne flows" in H II regions (e.g.
Alvarez et al. 2006). In this paper, we systematically explore
isothermal central voids in self-similar expansion and present
various forms of possible ISSV solutions. With gas self-gravity and
pressure, our model represents a fairly general theory to describe
the dynamic evolution of isothermal voids in astrophysical settings
on various spatial and temporal scales.



This paper is structured as follows. Section 1 is an introduction
for background information, motivation and astrophysical voids on
various scales. Section 2 presents the model formulation for
isothermal self-similar hydrodynamics, including the self-similar
transformation, analytic asymptotic solutions and isothermal shock
conditions. Section 3 explores all kinds of spherical ISSV solutions
constructed by the phase diagram matching method with extensions of
the so-called ``phase net". In Section 4, we demonstrate the
importance of the gas self-gravity, propose the physics on void edge
and then give several specific examples that the ISSV solutions are
applicable, especially in the contexts of PNe and interstellar
bubbles. Conclusions are summarized in Section 5. Technical details
are contained in Appendices A and B.

\section[]{Hydrodynamic Model Formulation}

We recount basic nonlinear Euler hydrodynamic equations in spherical
polar coordinates $(r,\theta,\varphi)$ with self-gravity and
isothermal pressure under the spherical symmetry.

\subsection[]{Nonlinear Euler Hydrodynamic Equations}

The mass conservation equations simply read
\begin{equation}
  \frac{\partial M}{\partial t}+u\frac{\partial M}{\partial r}=0\ ,\\
  \qquad\frac{\partial M}{\partial r}=4\pi r^2\rho\ ,\label{MCE}
\end{equation}
where $u$ is the radial flow speed, $M(r,\ t)$ is the enclosed mass
within radius $r$ at time $t$ and $\rho(r,\ t)$ is the mass density.
The differential form equivalent to continuity equation (\ref{MCE})
is
\begin{equation}
  \frac{\partial \rho}{\partial t}+\frac{1}{r^2}
  \frac{\partial}{\partial r}(r^2\rho
  u)=0\ .\label{CE}
\end{equation}
For an isothermal gas, the radial momentum equation is
\begin{equation}
  \frac{\partial u}{\partial t}+u\frac{\partial u}{\partial r}
  =-\frac{a^2}{\rho}\frac{\partial \rho}{\partial
  r}-\frac{GM}{r^2}\ ,\label{ME}
\end{equation}
where $G\equiv6.67\times10^{-8}$ dyn cm$^2$ g$^{-2}$ is the
gravitational constant, $a\equiv(p/\rho)^{1/2}=(k_{\rm
B}T/m)^{1/2}$ is the isothermal sound speed and $p$ is the gas
pressure, $k_{\rm B}\equiv 1.38\times 10^{-16}\hbox{ erg K}^{-1}$
is Boltzmann's constant, $T$ is the constant gas temperature
throughout and $m$ is the mean particle mass.
%
%
Meyer (1997) and Chevalier (1997a) ignored the gas self-gravity in
the momentum equation.

A simple dimensional analysis for equations $(\ref{MCE})-(\ref{ME})$
gives an independent dimensionless similarity variable
\begin{equation}
   x={r}/{(at)}\ ,\label{STx}
\end{equation}
involving the isothermal sound speed $a$. The consistent similarity
transformation is then
\begin{multline}\label{ST}
   \rho(r,\ t)=\alpha(x)/(4\pi G t^2)\ ,\\
   M(r,\ t)=a^3tm(x)/G\ ,\qquad
   u(r,\ t)=av(x)\ ,\\
\end{multline}
where $\alpha(x)$, $m(x)$, $v(x)$ are the dimensionless reduced
variables corresponding to mass density $\rho(r,t)$, enclosed mass
$M(r,t)$ and radial flow speed $u(r,t)$, respectively. These reduced
variables depend only on $x$ (Shu 1977;
Hunter 1977; Whitworth \& Summers 1985; Tsai \& Hsu 1995;
Shu et al 2002; Shen \& Lou 2004; Lou \& Shen 2004; Bian \& Lou
2005). Meyer (1997) adopts a different self-similar transformation
by writing $\rho=\bar{\rho}(x)/r^2$. By equation (\ref{STx})
above, we know that in Meyer (1997), $\bar{\rho}(x)$ is exactly
equal to $x^2\alpha(x) a^2/(4\pi G)$ here. So the similarity
transformation of Meyer (1997) is equivalent to similarity
transformation (\ref{ST}) here but without the self-gravity.
Further analysis will show that transformation (\ref{ST}) can
satisfy the void boundary expansion requirement automatically (see
Section 4.2.1).

With self-similar transformation (\ref{STx}) and (\ref{ST}),
equation (\ref{MCE}) yields two ordinary differential equations
(ODEs)
\begin{equation}
m+(v-x)\frac{dm}{dx}=0\ ,\qquad\qquad \frac{dm}{dx}=x^2\alpha\
.\label{enclosedmassd}
\end{equation}
The derivative term $dm/dx$ can be eliminated from these two
relations in equation (\ref{enclosedmassd}) to give
\begin{equation}
m(x)=x^2\alpha(x-v)\ .\label{enclosedmass}
\end{equation}
A nonnegative mass $m(x)$ corresponds to $x-v>0$. In figure
displays of $-v(x)$ versus $x$ profiles, the lower left portion to
the line $v-x=0$ is thus unphysical. We refer to the line $v-x=0$
as the ``Zero Mass Line" (ZML); inside of $x=v$, there should be
no mass, corresponding to a void.

Relation (\ref{enclosedmass}) plus transformation (\ref{STx}) and
(\ref{ST}) lead to two coupled ODEs from equations (\ref{CE}) and
(\ref{ME}), namely
\begin{equation}
\left[(x-v)^2-1\right]\frac{dv}{dx}=\left[\alpha(x-v)
-\frac{2}{x}\right](x-v)\ ,\label{ODE1}
\end{equation}
\begin{equation}
\left[(x-v)^2-1\right]\frac{1}{\alpha}\frac{d\alpha}{dx}
=\left[\alpha-\frac{2}{x}(x-v)\right](x-v)\ \label{ODE2}
\end{equation}
(Shu 1977).
ODEs (\ref{ODE1}) and (\ref{ODE2}) differ from eqs (3) and (4) of
Chevalier (1997a) by including the self-gravity effect.

For ODEs (\ref{ODE1}) and (\ref{ODE2}), the singularity at
$(x-v)^2=1$ corresponds to two parallel straight lines in the
diagram of $-v(x)$ versus $x$, representing the isothermal sonic
critical lines (SCL) (e.g. Shu 1977;
Whitworth \& Summers 1985; Tsai \& Hsu 1995; Shu et al. 2002;
Lou \& Shen 2004; Bian \& Lou 2005). As $x-v=-1<0$ is unphysical for
a negative mass, we have the SCL characterized by
\begin{equation}
\qquad v=x-1\ ,\\ \alpha={2}/{x}\ .\label{SCL}
\end{equation}
%
Two important global analytic solutions of nonlinear ODEs
(\ref{ODE1}) and (\ref{ODE2}) are
the static singular isothermal sphere (SIS; e.g. Shu 1977)
\begin{equation}
v=0\ ,\qquad\qquad \alpha=\frac{2}{x^2}\ ,\qquad\qquad m=2x\
,\label{SIS}
\end{equation}
and non-relativistic Einstein-de Sitter expansion solution for an
isothermal gas
\begin{equation}\label{EinsteinDSitter}
v=\frac{2}{3}x\ ,\qquad \alpha=\frac{2}{3}\ ,\qquad
m=\frac{2}{9}x^3\
\end{equation}
(e.g. Whitworth \& Summers 1985; Shu et al. 2002).

Let $x_*$ denote the value of $x$ at a sonic point on SCL. A Taylor
series expansion of $v(x)$ and $\alpha(x)$ in the vicinity of $x_*$
shows that solutions crossing the SCL smoothly at $x_*$ have the
form of either
\begin{equation}\label{deri1}
\begin{aligned}
  -v=(1-x_*)+\bigg({1\over x_*}-1\bigg)(x-x_*)+\cdots\ , \\
  \alpha={2\over x_*}-{2\over x_*}\bigg({3\over x_*}
  -1\bigg)(x-x_*)+\cdots\ ,
\end{aligned}
\end{equation}
  or
\begin{equation}\label{deri2}
\begin{aligned}
  -v=(1-x_*)-{1\over x_*}(x-x_*)+\cdots\ ,  \\
  \alpha={2\over x_*}-{2\over x_*}(x-x_*)+\cdots\ ,
\end{aligned}
\end{equation}
(e.g. Shu 1977;
Whitworth \& Summers 1985; Tsai \& Hsu 1995; Bian \& Lou 2005;
see Appendix A of Lou \& Shen 2004 for higher-order derivatives).
Thus eigensolutions crossing the SCL smoothly are uniquely
determined by the value of $x_*$. Eigensolutions of type 1 and 2 are
defined by equations (\ref{deri1}) and (\ref{deri2}). Physically,
equations (\ref{deri1}) and (\ref{deri2}) describe how the gas
behaves as it flows from subsonic to supersonic regimes across the
SCL in the local comoving framework.

An important numerical solution to ODEs (\ref{ODE1}) and
(\ref{ODE2}) is the Larson-Penston (LP) solution (Larson 1969a;
Larson 1969b; Penston 1969). This solution have an asymptotic
behaviour $v\rightarrow 2x/3$ and $\alpha\rightarrow1.67$ as
$x\rightarrow0^+$. And the LP solution is also an eigensolution of
type 2 (equation \ref{deri2}) and passes through the SCL smoothly at
$x_*=2.33$.

When $x\rightarrow+\infty$, either at large radii or at very early
time, solutions to two nonlinear ODEs (\ref{ODE1}) and (\ref{ODE2})
have asymptotic behaviours
\begin{multline}\label{largeasymp}
    v=V+\frac{2-A}{x}+\frac{V}{x^2}
    +\frac{(A/6-1)(A-2)+2V^2/3}{x^3}+\cdots\ ,\\
    \alpha=\frac{A}{x^2}+\frac{A(2-A)}{2x^4}
    +\frac{(4-A)VA}{3x^5}+\cdots\ ,\\
    m=Ax-AV+\frac{A(2-A)}{2x}
    +\frac{A(A-4)V}{6x^2}+\cdots\ ,\\
\end{multline}
where $V$ and $A$ are two integration constants (e.g. Whitworth \&
Summers 1985; Lou \& Shen 2004), referred to as the reduced velocity
and mass parameters, respectively.
The non-relativisitic Einstein-de Sitter expansion solution does not
follow this asymptotic solution (\ref{largeasymp}) at large $x$.
Chevalier (1997a) also presented asymptotic solutions of $v(x)$ and
$\alpha(x)$ at large $x$. Case $A=1$ in our solution
(\ref{largeasymp}) should correspond to asymptotic behaviours of
$v(x)$ and $\alpha(x)$ in Chevalier's model (1997a; see his
equations 5 and 6). However, the coefficient of $x^{-4}$ term for
$\alpha(x)$ in his model is $1$ while in our model, it is $1/2$; and
coefficient of $x^{-1}$ term for $v(x)$ in his model is $2$ while in
our model, it is $2-A$. These differences arise by dropping the
gravity in Chevalier (1997a). Physically in our model, the gas has a
slower outgoing radial velocity and the density decreases more
rapidly than that of Chevalier (1997a), because when the
self-gravity is included, the gas tends to accumulate around the
centre.
Counterpart solutions to equations (\ref{SIS})$-$(\ref{largeasymp})
can also be generalized for conventional and general polytropic
gases (Lou \& Wang 2006; Wang \& Lou 2007; Lou \& Cao 2008; Wang \&
Lou 2008; Hu \& Lou 2008).





\subsection[]{Isothermal Shock Jump Conditions}

For an isothermal fluid, the heating and driving of a cloud or a
progenitor star (such as a sudden explosion of a star) at $t=0$ will
compress the surrounding gas and give rise to outgoing shocks (e.g.
Tsai \& Hsu 1995). In the isothermal approximation, the mass and
momentum should be conserved across a shock front in the shock
comoving framework of reference
\begin{equation}\label{shockcm}
\rho_d(u_d-u_s)=\rho_u(u_u-u_s)\ ,
\end{equation}
\begin{equation}\label{shockcmom}
 a_d^2\rho_d+\rho_du_d(u_d-u_s)=a_u^2\rho_u+\rho_uu_u(u_u-u_s)\ ,
\end{equation}
where subscripts $d$ and $u$ denote the downstream and upstream
sides of a shock, respectively (e.g. Courant \& Friedricks 1976;
Spitzer 1978; Dyson \& Williams 1997; Shen \& Lou 2004; Bian \& Lou
2005). Physically, we have $u_s=a_dx_{sd}=a_ux_{su}=r_s/t$ as the
outgoing speed of a shock with $r_s$ being the shock radius.
Conditions (\ref{shock1}) and (\ref{shock2}) below in terms of
self-similar variables are derived from conditions (\ref{shockcm})
and (\ref{shockcmom}) by using the reduced variables $v(x)$, $x$ and
$\alpha(x)$ and the isothermal sound speed ratio $\tau\equiv
a_d/a_u=x_{su}/x_{sd}$,
\begin{equation}\label{shock1}
  \alpha_d/\alpha_u=(v_u-x_{su})/[\tau(v_d-x_{sd})]\ ,
\end{equation}
\begin{equation}\label{shock2}
  v_d-x_{sd}-\tau (v_u-x_{su})=(\tau v_d-v_u)(v_u-x_{su})(v_d-x_{sd})\ .
\end{equation}
Consequently, we have $\tau=(T_d/T_u)^{1/2}$ with $T$ being the gas
temperature. Physics requires $T_d\ge T_u$ leading to $\tau\ge 1$.
For $\tau=1$, conditions (\ref{shock1}) and (\ref{shock2}) reduce to
\begin{equation}\label{shock1tau1}
  \alpha_d/\alpha_u=(v_u-x_{s})/(v_d-x_{s})\ ,
\end{equation}
\begin{equation}\label{shock1tau2}
  (v_u-x_s)(v_d-x_s)=1\ ,
\end{equation}
where $x_s=x_u=x_d$ is the reduced shock location or speed (e.g.
Tsai \& Hsu 1995; Chevalier 1997a; Shu et al. 2002; Shen \& Lou
2004; Bian \& Lou 2005).


\section[]{Isothermal Self-Similar Voids}


Various similarity solutions can be constructed to describe outflows
(e.g. winds and breezes), inflows (e.g. contractions and
accretions), static outer envelope and so forth. These solutions
will be presented below in order, as they are useful to construct
ISSV solutions.

\subsection[]{Several Relevant Self-Similar Solutions}

We first show some valuable similarity solutions in reference to
earlier results of Shu (1977) and  Lou \& Shen (2004). These
solutions behave differently as $x\rightarrow+\infty$ for various
combinations of parameters $V$ and $A$ in asymptotic solution
(\ref{largeasymp}).

\subsubsection[]{CSWCP and EWCS Solutions of Shu (1977)}

\begin{figure}
\centering
\includegraphics[width=0.50\textwidth]{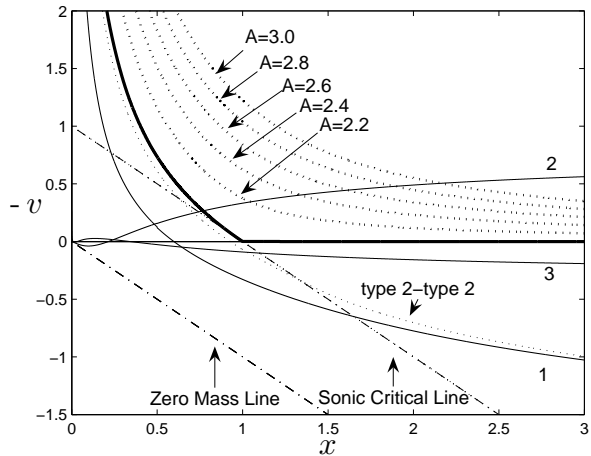}
\caption{
Isothermal self-similar solutions of Shu (1977): $-v(x)$ versus $x$
profile of EWCS (heavy solid curve) and a sequence of five CSWCP
solutions (dotted curves) with mass parameter $A$ values marked
along these five curves ($A=2.2$ to 3.0). Four solutions smoothly
crossing the SCL (dash-dotted line to the upper right) twice of Lou
\& Shen (2004): three light solid curves labelled by numerals `1',
`2' and `3' are type 2-type 1 solutions with their smaller crossing
points \textbf{$x_*(1)$} matching the type 2 derivative and the
larger crossing points \textbf{$x_*(2)$} fitting the type 1
derivative. The light dotted curve labelled by `type 2-type 2' is
the unique type 2-type 2 solution with both crossing points
following the type 2 derivative (see Table 1 for relevant
parameters). The dash-dotted line to the lower left is the ZML of
$x-v=0$.}\label{fig:CSWCP_EWCS_EECC}
\end{figure}

Shu (1977) presents a class of solutions: collapse solutions without
critical point (CSWCP) and expansion-wave collapse solution (EWCS)
(see Fig. \ref{fig:CSWCP_EWCS_EECC}).

The CSWCP solutions (light dotted curves in Fig.
\ref{fig:CSWCP_EWCS_EECC}) have asymptotic behaviours of $V=0$ and
$A>2$ according to solution (\ref{largeasymp}) at large $x$, and
describe the central free-fall collapse of gas clouds with
contracting outer envelopes of vanishing velocities at large radii.

The EWCS solution (the heavy solid curve in Fig.
\ref{fig:CSWCP_EWCS_EECC}) is obtained with
$A\rightarrow 2^+$ and
$V=0$ in solution (\ref{largeasymp}). This solution is tangent to
the SCL at $x_*=1$ and has an outer static SIS envelope (solution
\ref{SIS}) and a free-fall collapse towards the centre (
Shu 1977; Lou \& Shen 2004; Bian \& Lou 2005); the central
collapsed region expands in a self-similar manner.

\subsubsection[]{Solutions smoothly crossing the SCL twice}

Lou \& Shen (2004) studied isothermal similarity solutions and
divided Class I similarity solutions, which follow free-fall
behaviours as $x\rightarrow 0^+$, into three subclasses according to
their behaviours at large $x$. Class Ia similarity solutions have
positive
$V$ at large $x$ (see solution \ref{largeasymp}), which describe a
cloud with an envelope expansion. Class Ia solutions are referred to
as `envelope expansion with core collapse' (EECC) solutions by Lou
\& Shen (2004). The case of $V=0$ corresponds to Class Ib solutions
and $V<0$ renders Class Ic solutions. The CSWCP solutions belong to
Class Ic and the EWCS solution belongs to Class Ib.

Lou \& Shen (2004) constructed four discrete solutions crossing the
SCL twice smoothly (i.e. satisfying equations \ref{deri1} and
\ref{deri2}) by applying the phase diagram matching scheme (Hunter
1977; Whitworth \& Summers 1985; see subsection 3.4 below).
These are the first four examples among an infinite number of
discrete solutions (see Lou \& Wang 2007).

Let $x_*(1)$ and $x_*(2)$ be the smaller (left) and larger (right)
cross points for each of the four solutions. As they are all Class I
solution, $x_*(1)$ is less than $1$ and the behaviours near $x_*(1)$
are determined by type 2 (as defined by equation \ref{deri2}) to
assure a negative derivative $d(-v)/dx$ at $x_*(1)$ along the SCL.
As in Lou \& Shen (2004), we name a solution that crosses the SCL
twice smoothly as `type2-type1' or `type2-type2' solution, which
corresponds the derivative types at $x_*(1)$ and $x_*(2)$. Three of
the four solutions of Lou \& Shen (2004) are `type2-type1' solutions
and we further name them as `type2-type1-1', `type2-type1-2' and
`type2-type1-3' (see Fig. \ref{fig:CSWCP_EWCS_EECC} and Table 1 for
more details).



\begin{table}
\tabcolsep 0pt\caption{Relevant parameters are summarized here for
the four solutions crossing the SCL twice shown in Figure
\ref{fig:CSWCP_EWCS_EECC}.
Three of them cross the SCL at $x_*(1)$ with the second kind of
derivative and at $x_*(2)$ with the first kind of derivative. In the
`Solution' column of this table, they are named as `type2-type1-1',
`type2-type1-2' and `type2-type1-3' corresponding to the curves
labelled by `1', `2' and `3', respectively. The fourth solution
crosses the SCL at $x_*(1)$ and $x_*(2)$ both with the second kind
of derivative and is named as `type2-type2' in the `Solution'
column, corresponding to the curve labelled by `type 2-type 2'. The
first and third type2-type1 solutions and type2-type2 solution all
belong to EECC solution (or Class Ia solution), while the
type2-type1-2 solution belongs to Class Ic solution. In Lou \& Shen
(2004), this type2-type2 EECC solution passes the SCL at
$x_*(1)\approx 0.632$ and $x_*(2)\approx 1.349$. We reproduce these
results.
However, we here adopt $x_*(1)\approx0.71$ and $x_*(2)\approx
1.287$, which are calculated with a higher
accuracy. }\vspace*{-10pt}
\begin{center}
\def\temptablewidth{0.45\textwidth}
{\rule{\temptablewidth}{1pt}}
\begin{tabular*}{\temptablewidth}{@{\extracolsep{\fill}}c|c|c|c|c|c}
Solution     & Class    & $x_*(1)$                &$x_*(2)$ & $V$  & $A$ \\
\hline
type2-type1-1& EECC(Ia) & $0.23$                & $1.65$  &  $1.8$  & $5.1$\\
type2-type1-2& Ic       & $2.5858\times 10^{-4}$& $0.743$ & $-0.77$ & $1.21$ \\
type2-type1-3& EECC(Ia) & $6\times 10^{-6}$     & $1.1$   &  $0.3$  & $2.4$ \\
type2-type2  & EECC(Ia) & $0.71$ & $1.287$ & $1.5$ & $4.7$
       \end{tabular*}
       {\rule{\temptablewidth}{1pt}}
       \end{center}
       \end{table}

\subsection[]{Isothermal Self-Similar Void (ISSV) Solutions}

Relation (\ref{enclosedmass}) indicates that the ZML $v-x=0$
separates the solution space into the upper-right physical part and
the lower-left unphysical part
in a $-v(x)$ versus $x$ presentation.

For a solution (with $x-v>0$) touching the ZML at $x_0$, then
$v(x_0)=x_0$ holds on and so does $m(x_0)=0$ there. Given the
definitions of $m(x)\equiv GM(r,\ t)/(a^3t)$
and $x\equiv r/(at)$, condition $m(x_0)=0$ indicates a spherical
isothermal gas whose enclosed mass $M(atx_0, t)$ vanishes, and thus
a central void expands at a constant radial speed $ax_0$.

The condition $v(x_0)=x_0$ marks a void boundary in expansion.
Physically, $x_0$ is the start point of a streamline as $v(x_0)=x_0$
indicates matters flowing outwards at a velocity of the boundary
expansion velocity.
However, a problem would arise when the gas density just outside the
void boundary is not zero. For an isothermal gas, the central vacuum
cannot resist an inward pressure from the outer gas across the void
boundary. We propose several mechanisms/scenarios to provide
sufficient energy and pressure to generate voids and maintain them
for a period of time without invalidating our `vacuum approximation'
for voids.
In Section 4, we present explanations and
quantitative calculations about these mechanisms.

Mathematically, to construct spherical isothermal self-similar voids
is to search for global solutions reaching the ZML at $x_0>0$.
If a solution touches the ZML at point $x_0$, then $x_0$ should be
the smallest point of the solution with gas.
Solutions with a negative mass are unphysical.
So if $v=x$ holds on at $x_0$, both $v(x)$ and $\alpha(x)$ should be
zero in the range $0<x<x_0$. \footnote{
The vacuum solution $v=0$ and $\alpha=0$ satisfies dimensional
equations (\ref{MCE})$-$(\ref{ME}); it is not an apparent solution
to reduced ODE (\ref{ODE1}) because a common factor $\alpha$ has
been taken out before we arrive at ODE (\ref{ODE1}). With this
physical understanding, we still regard $v=0$ and $\alpha=0$ as a
solution to ODEs (\ref{ODE1}) and (\ref{ODE2}). }

Before exploring isothermal self-similar void (ISSV) solutions of
spherical symmetry, we note a few properties of such solutions.
Nonlinear ODEs (\ref{ODE1}) and (\ref{ODE2}) give the following two
first derivatives at $x_0$
\begin{equation}\label{voidedge}
  \frac{dv}{dx}\bigg|_{x_0}=0\ ,
  \qquad\qquad\frac{d\alpha}{dx}\bigg|_{x_0}=0\ .
\end{equation}
So in the $-v(x)$ versus $x$ and $\alpha(x)$ versus $x$
presentations, the right tangents to these solution curves
at $x_0$ are horizontal.

For spherical isothermal self-similar dynamics, we can show that
across a void boundary, $\alpha(x)$ must jump from $0$ to a nonzero
value (see Appendix A).
Voids must be bounded by relatively dense shells with a density
jump.
We propose that the height of such a density jump may indicate the
energy to generate and maintain such a void. Energetic processes of
short timescales include supernovae involving rebound shock waves or
neutrino emissions and driving processes of long timescales include
powerful stellar winds (see Section 4 for more details).
In reality, we do not expect an absolute vacuum inside a void.
Regions of significantly lower density in gas clouds are usually
identified as voids.

For $x\rightarrow+\infty$, the physical requirement of finite mass
density and flow velocity can be met for $\alpha(x)$ and $v(x)$ by
asymptotic solution (\ref{largeasymp}).

ISSV solutions need to cross the SCL in the $-v(x)$ versus $x$
profiles as they start at the ZML and tend to a horizontal line at
large $x$ with a constant $V$. Given conditions (\ref{deri1}) and
(\ref{deri2}), ISSV solutions can be divided into two subtypes:
crossing the SCL smoothly without shocks, which will be referred to
as type $\mathcal{X}$, and crossing the SCL via shocks, which will
be referred to as type $\mathcal{Z}$ and can be further subdivided
into types $\mathcal{Z}_{\rm I}$ and $\mathcal{Z}_{II}$ as explained
presently.

\subsection[]{Type $\mathcal{X}$ ISSV
Solutions without Shocks}

As analyzed in Section 3.2, type $\mathcal{X}$ ISSV solutions cross
the SCL smoothly. Let $x_*$ denote the cross point on the SCL. We
have $v(x_*)=x_*-1$ and $\alpha(x_*)=2/x_*$ by equation (\ref{SCL}).
Conditions (\ref{deri1}) and (\ref{deri2}) give the eigen solutions
for the first derivatives $v'(x)$ and $\alpha'(x)$ at $x_*$ as
either type 1 (equation {\ref{deri1}) or type 2 (equation
\ref{deri2}). Given $x_*$ and the type of eigen-derivative at $x_*$,
all the necessary initial conditions [$v(x_*)$, $\alpha(x_*)$,
$v'(x_*)$, $\alpha'(x_*)$] are available for integrating nonlinear
ODEs (\ref{ODE1}) and (\ref{ODE2}) in both directions.


We use $x_0$, $\alpha_0\equiv\alpha(x_0)$, $x_*$ and the types of
eigen-derivative at $x_*$ to construct type $\mathcal{X}$ ISSV
solutions. While $x_0$ and $\alpha_0$ are the key parameters for the
void expansion speed and the density of the shell around the void
edge, we use $x_*$ to obtain type $\mathcal{X}$ ISSV solutions as
$x_*$ parameter can be readily varied to explore all
type $\mathcal{X}$ ISSV solutions.

\subsubsection[]{Type $\mathcal{X}_{\rm I}$ ISSV
Solutions: Voids with Sharp Edge, Smooth Envelope and Type 1
Derivative on the SCL}

Type $\mathcal{X}_{\rm I}$ solutions cross the SCL smoothly and
follow the Type 1 derivative at $x_*$ on the SCL. By equation
(\ref{deri1}), the first derivative $d(-v)/dx$ is positive for
$0<x_*<1$ and negative for $x_*>1$. This allows the $x_*$ of type
$\mathcal{X}_{\rm I}$ solutions to run from $0$ to $+\infty$ when
behaviours of these solutions in the inner regions are ignored
temporarily. When the existence of $x_0$ is required for
constructing isothermal central voids, the range of $x_*$ along the
SCL is then restricted.

Given $x_*$ on the SCL and using equations (\ref{SCL}) and
(\ref{deri1}), we obtain the initial condition $v(x_*),\
\alpha(x_*),\ v'(x_*)$, $\alpha'(x_*)$ to integrate ODEs
(\ref{ODE1}) and (\ref{ODE2}) from $x_*$ in both directions. If an
integration towards small $x$ can touch the ZML at a $x_0>0$ and
an integration towards $+\infty$ exists, then a type
$\mathcal{X}_{\rm I}$ ISSV solution is constructed.

We now list five important numerals: $x_1\approx 0.743$, $x_2=1$,
$x_3\approx 1.1$, $x_4\approx 1.65$ and $x_5=3$. They are the cross
points at the SCL of Lou \& Shen type2-type1-2, SIS (see equation
11), Lou \& Shen type2-type1-3, Lou \& Shen type2-type1-1,
Einstein-de Sitter solution, respectively. All these solutions have
type 1 derivatives near their cross points on the SCL. However,
their behaviours differ substantially as $x\rightarrow 0^+$. The
three solutions of Lou \& Shen approach central free-fall collapses
with a constant reduced core mass $m_0$ as the central mass
accretion rate; while SIS and Einstein-de Sitter solution have
vanishing velocities as $x\rightarrow 0^+$.

Numerical computations show that type $\mathcal{X}_{\rm I}$ ISSV
solutions exist when $x_*$ falls into four intervals out of six
intervals along $x>0$ axis divided by the five numerals above. The
four intervals are $0<x_*<x_1\approx 0.743$, $x_1\approx 0.743
<x_*<x_2=1$, $x_2=1<x_*<x_3\approx1.1$ and
$x_3\approx1.1<x_*<x_4\approx1.65$. No type $\mathcal{X}_{\rm I}$
exist with $x_*$ in intervals $x_4\approx1.65<x_*<x_5=3$ and
$x_*>x_5=3$, because integrations from $x_*$ in these two intervals
towards $+0$ or $+\infty$ must halt when they encounter the SCL
again, respectively. The six regions mentioned above are named as
conditions I, II, III, IV, V and VI, respectively.
Figure \ref{fig:typealphaI} illustrates several typical type
$\mathcal{X}_{\rm I}$ ISSV solutions with their $x_*$ in different
regions. The relevant solution parameters are summarized in Table 2.

\begin{figure}
\centering
\includegraphics[width=0.50\textwidth]{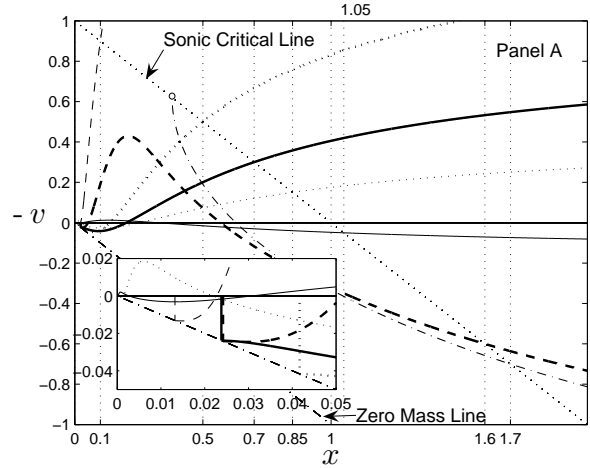}
\includegraphics[width=0.50\textwidth]{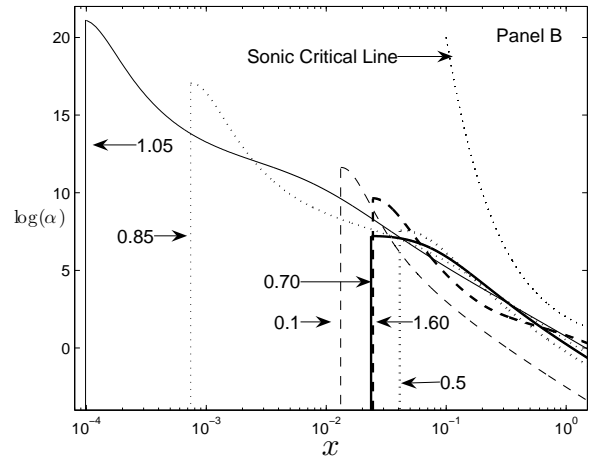}
\caption{Typical type $\mathcal{X}_{\rm I}$ ISSV solutions. Panel A
in linear scales shows $-v(x)$ versus $x$ curves of several type
$\mathcal{X}_{\rm I}$ solutions. The sonic critical point $x_*$ of
each curve is noted along the $x-$axis. The dash-dotted curve which
crosses the SCL at $x_*=1.7$ smoothly and encounters the line again
at $\sim 0.4$ shows a typical behaviour when
$x_4\approx1.65<x_*<x_5=3$. The inset shows the enlarged portions of
these ISSV solution curves near $x\rightarrow 0^+$. The dash-dotted
lines in both panel A and inset are the ZML. Panel B shows the
$\log\alpha$ versus $x$ curves of the same solutions and the curves
in panel A, inset and panel B with the same line type (light and
heavy solid, dotted and dash) correspond to the same type
$\mathcal{X}_{\rm I}$ solutions. They are distinguished by their
$x_*$ values. Each of these curves jumps from a zero value (left) to
a nonzero value (right) at $x_0$, indicating a void in the region of
$x<x_0$. }\label{fig:typealphaI}
\end{figure}

\begin{table}
\tabcolsep 0pt \caption{Parameters of several typical type
$\mathcal{X}_{\rm I}$ ISSV solutions. 
Eight solutions under four different conditions discussed in Section
3.3.1 are tabulated here.}\vspace*{-12pt}
\begin{center}
\def\temptablewidth{0.45\textwidth}
{\rule{\temptablewidth}{1pt}}
\begin{tabular*}{\temptablewidth}{@{\extracolsep{\fill}}c|c|cc|cc}
Condition&$x_*$  &  $x_0$              &$\alpha_0$        & $V$   & $A$   \\
\hline
I        &$0.10$ &  $0.0132$           &$1.11\times10^5$  &$-3.51$&$0.070$\\
I        &$0.50$ &  $0.0417$           &$1.92\times10^3$  &$-1.57$&$0.649$\\
I        &$0.70$ &  $0.0238$           &$1.35\times10^3$  &$-0.91$&$1.10 $\\
         &       &                     &                  &       &       \\
II       &$0.75$ &  $7.5\times10^{-5}$ &$1.5\times10^{10}$&$-0.75$&$1.23 $\\
II       &$0.85$ &  $7.5\times10^{-4}$ &$2.5\times10^7$   &$-0.46$&$1.50 $\\
II       &$0.95$ &  $5.6\times10^{-4}$ &$6.9\times10^6$   &$-0.15$&$1.83 $\\
         &       &                     &                  &       &       \\
III      &$1.05$ &  $9.2\times10^{-5}$ &$1.5\times10^9$   &$+0.14$&$2.17 $\\
         &       &                     &                  &       &       \\
IV       &$1.60$ & $0.0242$            &$1.5\times10^4$   &$+1.71$&
$4.77$
\end{tabular*}
{\rule{\temptablewidth}{1pt}}
\end{center}
\end{table}

\subsubsection[]{Type $\mathcal{X}_{\rm II}$ Void Solutions: Voids with Sharp Edges,
Smooth Envelope and Type 2 Derivative at the SCL}

Type $\mathcal{X}_{\rm II}$ solutions cross the SCL smoothly and
have the type 2 derivative at $x_*$. By condition (\ref{deri2}), the
first derivative $d(-v)/dx$ is negative for $x_*>0$. Because
$x_0<x_*$, $x_*$ of type $\mathcal{X}_{\rm II}$ solutions must be
larger than $1$ to assure behaviours of solution (\ref{largeasymp})
at large $x$ (note that type 2 derivative is used to obtain a
free-fall behaviour around the centre). Similar to the approach to
investigate type $\mathcal{X}_{\rm I}$ ISSV solutions, we now list
four important numerals: $x_1'=1$, $x_2'\approx1.287$, $x_3'=1.50$
and $x_4'\approx2.33$. Here, $x_2'\approx1.287$ is the right cross
point on the SCL of the type2-type2 solution of Lou \& Shen (2004)
and $x_4'\approx 2.33$ is the cross point of the LP solution on the
SCL. These two solutions both follow behaviours of type 2 derivative
near their cross points. However, their behaviours differ as
$x\rightarrow 0^+$. Lou \& Shen type 2-type 2 EECC solution has a
central free-fall collapse with a constant reduced core mass $m_0$
for the central mass accretion rate; while LP solution has vanishing
velocity and mass as $x\rightarrow 0^+$. And $x_3'=1.5$ is the
critical point where the second-order type 2 derivative diverges
(see Appendix A of Lou \& Shen 2004).

The four points given above subdivide the $x\geq1$ portion of the
$x-$axis into four intervals: $[x_1'=1, x_2'\approx 1.287]$,
$[x_2'\approx1.287, x_3'=1.5]$, $[x_3'=1.5, x_4'\approx 2.34]$ and
$[x_4'\approx 2.34, +\infty)$ which are referred to as condition I',
II', III', and IV', respectively. Similar to subsection 3.3.1, we
choose a $x_*$ and integrate from $x_*$ in both directions.
Numerical calculations indicate that type $\mathcal{X}_{\rm II}$
void solutions only exist when their $x_*$ falls under conditions
II' or IV'.

We show four typical type $\mathcal{X}_{\rm II}$ ISSV solutions in
Figure \ref{fig:typealphaII} with relevant parameters summarized in
Table 3.

\begin{figure}
\centering
\includegraphics[width=0.50\textwidth]{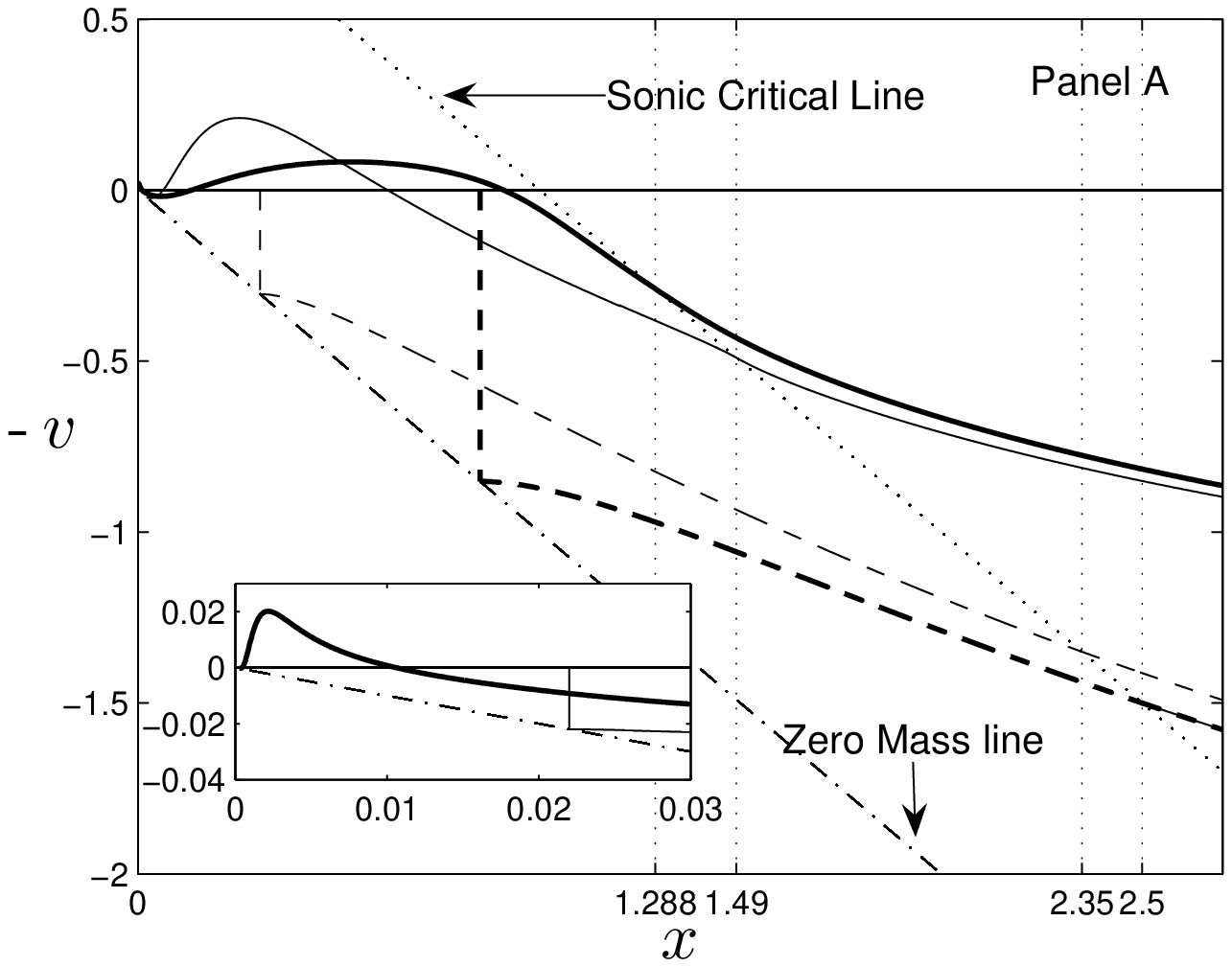}
\includegraphics[width=0.5\textwidth]{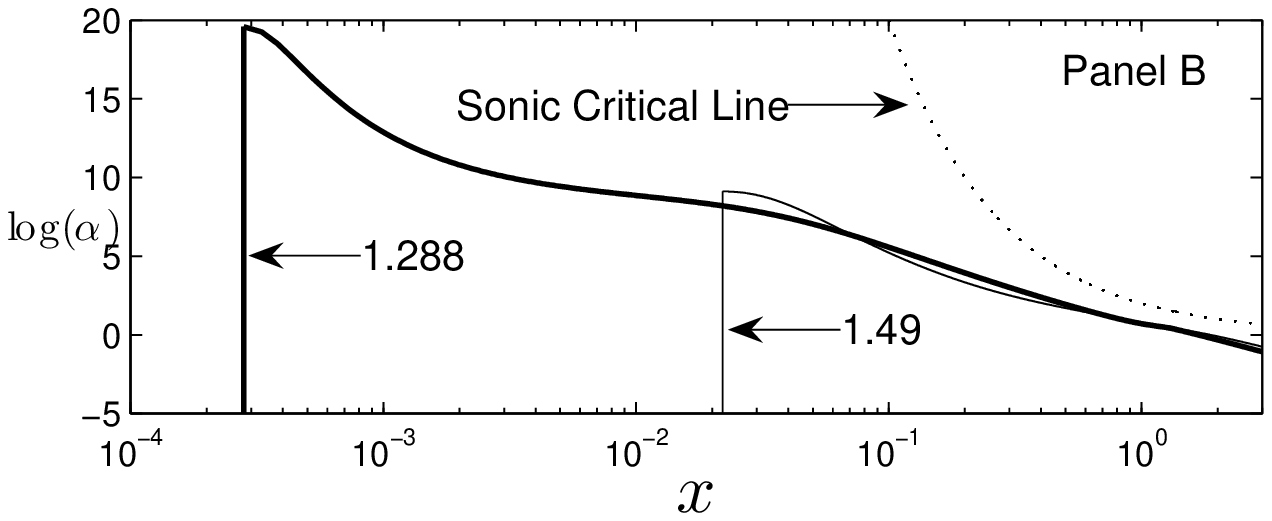}
\includegraphics[width=0.5\textwidth]{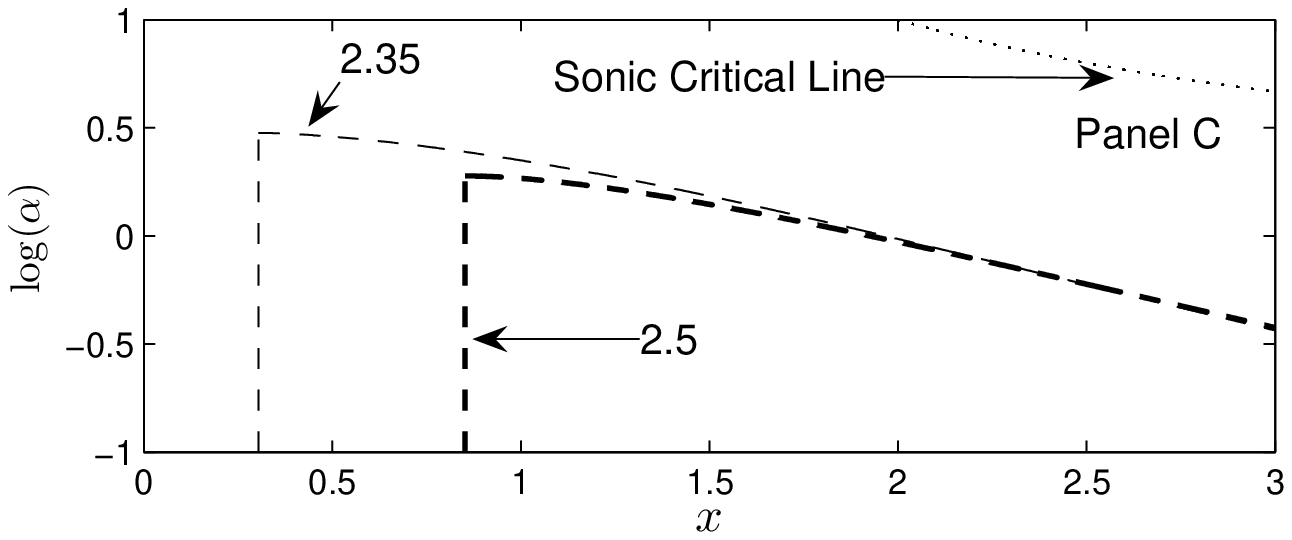}
\caption{Four typical type $\mathcal{X}_{\rm II}$ ISSV solutions in
$-v(x)$ versus $x$ profile. Values of their $x_*$ are marked by the
grid lines of $x-$axis in panel A. The inner box presents details of
two curves near $x\rightarrow 0^+$ in the same line type as
corresponding curves in Panel A. Panel B and C give the four
solutions in $\log\alpha$ versus $x$ profile. Panel B presents two
solution curves of II' condition marked with $x_*=1.288$ (heavy
solid line) and $x_*=1.49$ (light solid line) respectively and panel
C shows two solution curves of IV' condition marked with $x_*=2.35$
(light dash line) and $x_*=2.5$ (heavy dash line) respectively. The
density jumps at $x_0$ are obviously much sharper in condition II'
than in condition IV' (compare the numbers along $y-$axes).
}\label{fig:typealphaII}
\end{figure}

\begin{table}
\tabcolsep 0pt \caption{Parameters of several typical type
$\mathcal{X}_{\rm II}$ ISSV solutions. Five solutions under two
conditions in Section 3.3.1 are listed below.} \vspace*{-12pt}
\begin{center}
\def\temptablewidth{0.45\textwidth}
{\rule{\temptablewidth}{1pt}}
\begin{tabular*}{\temptablewidth}{@{\extracolsep{\fill}}c|c|cc|cc}
Condition&$x_*$&  $x_0$              &$\alpha_0$    &$V$&$A$\\
\hline
II' &$1.288$&$2.8\times10^{-4}$&$4\times10^8$  &$1.5$ & $4.7$\\
II' & $1.49$&$0.022$           &$9.1\times10^3$&$4.5$ &  $18$\\
    &       &                  &               &      &      \\
IV' & $2.35$&$0.304$           &$1.61$         &$3.30$&$8.54$\\
IV' & $2.5$ &$0.852$           &$1.32$         &$3.46$&$8.92$\\
IV' & $2.58$&$1.007$           &$1.23$         &$9.48$&$3.54$
\end{tabular*}
{\rule{\temptablewidth}{1pt}}
\end{center}
\end{table}

\subsubsection[]{Interpretations for Type $\mathcal{X}$ ISSV Solutions}

We have explored all possible type $\mathcal{X}$ ISSV solutions. Now
we offer interpretations for these solutions. In preceding sections,
we used parameter $x_*$ on the SCL as the key parameter to construct
type $\mathcal{X}$ ISSV solutions. This method assures us not to
miss any possible ISSV solutions. However, for ISSV solutions, $x_0$
and $\alpha_0$ are the most direct parameters describing properties
of central void expansions.

\begin{figure}
\centering
\includegraphics[width=0.50\textwidth]{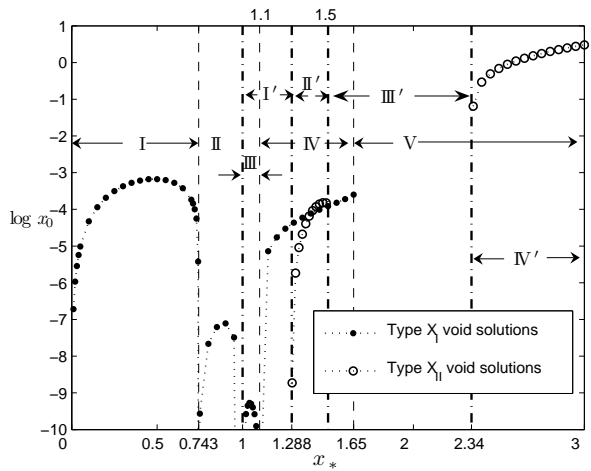}
\caption{The relation between isothermal void boundary $x_0$ and the
sonic point $x_*$ at SCL. The `solid circle curve' and the `open
circle curve' are the $\log(x_0)$ versus $x_*$ curves of type
$\mathcal{X}_{\rm I}$ and type $\mathcal{X}_{\rm II}$ void
solutions. The key numerals dividing the positive real $x-$axis into
several ranges are indicated by the dash vertical lines with
numerals marked along or at $x-$axis. In regions I, II, III, and IV,
we have four branches for type $\mathcal{X}_{\rm I}$ void solutions.
In regions II' and IV', we have two branches for type
$\mathcal{X}_{\rm II}$ ISSV solutions.}\label{fig:typealpha_x_*_x_0}
\end{figure}

The void edge $x_0$ is the reduced radius in the self-similar
description and $ax_0$ is the expansion speed of the void boundary.
We adjust the $x_*$ value at the SCL to search for void solutions
and there exists a certain relationship between $x_*$ and $x_0$ as
expected. Figure \ref{fig:typealpha_x_*_x_0} shows these
relationships for types $\mathcal{X}_{\rm I}$ and $\mathcal{X}_{\rm
II}$ ISSV solutions.

Under conditions I, II, III and given $x_*$, the corresponding $x_0$
runs from $0$ to a maximum value and then back to $0$. The maximum
values of conditions I, II and III are $x_0=0.042$,
$x_0=8\times10^{-4}$ and $x_0=0.9\times10^{-4}$ which correspond to
$x_*\approx 0.5$, $x_*\approx 0.89$ and $x_*\approx 1.04$,
respectivly. For $x_*$ under conditions IV or II', the corresponding
$x_0$ increases with $x_*$ and the intervals of $x_0$ are $[0^{+},
0.027]$ and $[0^{+}, 0.022]$, respectively. Under condition IV',
$x_0$ increases with $x_*$ from $[0^{+}, +\infty)$ monotonically.
However, when $x_*$ is under condition IV', the corresponding $x_0$
is usually very large (at least 100 times larger) compared to other
type $\mathcal{X}$ ISSV solutions unless $x_*$ is near $2.34$.
Physically under condition IV', isothermal voids usually expand at a
relatively high speed. Based on the value and meaning of $x_0$, type
$\mathcal{X}$ ISSV solutions can be divided into two classes: rapid
and slow ISSV expansion solutions. All type $\mathcal{X}_{\rm I}$
ISSV solutions and type $\mathcal{X}_{\rm II}$ ISSV solutions under
condition II' belong to slow void expansion solutions. Type
$\mathcal{X}_{\rm II}$ ISSV solutions under condition IV' belong to
rapid void expansion solutions.

Parameter $\alpha_0$ represents the reduced density $\alpha$ at the
void boundary. Figures \ref{fig:typealphaI} and
\ref{fig:typealphaII} and Tables 2 and 3 clearly indicate that
isothermal voids, described by type $\mathcal{X}$ ISSV solutions,
are all surrounded by dense mass shells and the gas density around
voids attenuates monotonically with increasing radius. So there are
classes of voids that evolve
with fairly sharp edges but without shocks. Nevertheless, sharp edge
around voids is not a general property of all void solutions.
Expanding voids with shocks can be surrounded by quasi-smooth edges
(never smooth edges as shown in Appendix A). In
Section 4, we will show that $\alpha_0$ is an important parameter
which may reveal the mechanism that generates and sustains a void.
Large $\alpha_0$ requires very energetic mechanisms against the high
inward pressure across the boundary. So type $\mathcal{X}$ voids may
be difficult to form because all type $\mathcal{X}$ voids have dense
boundaries.

Figures \ref{fig:typealphaI} and \ref{fig:typealphaII} and Tables 2
and 3 clearly show that all type $\mathcal{X}_{\rm II}$ void
solutions and type $\mathcal{X}_{\rm I}$ ISSV solutions under
conditions III and IV describe isothermal voids surrounded by gas
envelopes
in expansion (i.e. velocity parameter $V$ at $x\rightarrow\infty$
are positive). Astrophysical void phenomena are usually coupled with
outflows (i.e. winds). Our ISSV solutions indicate that rapidly
expanding voids must be surrounded by outflows.

Type $\mathcal{X}_{\rm I}$ ISSV solutions under conditions I and II
describe voids surrounded by contracting envelopes, although under
these two conditions the voids expand very slowly ($\leq 0.042a$,
see subsection 3.3.1) and are surrounded by very dense shells (see
Table 2).

Outflows and inflows are possible as indicated by type
$\mathcal{X}$ ISSV solutions, but no static shell is found in type
$\mathcal{X}$ ISSV solutions. In the following section, we show
voids with shocks being surrounded by static envelopes.

A clarification deems appropriate here that division of the $x>0$
axis in subsection 3.3.1 is actually not precise. In subsection
3.3.1, we divide $(0, +\infty)$ into six intervals by five points
$x_1$ to $x_5$ and $x_1$, $x_3$ and $x_4$ are the cross points at
the SCL of Lou \& Shen type2-type1 solution. We note that they are
only the first three examples of an infinite number of discrete
solutions
that cross the SCL smoothly twice via type 2
derivative first at a smaller $x$ and then type 1 derivative at a
larger $x$
(Lou \& Shen 2004). Our numerical computations show
that the fourth type2-type1 solution will pass the SCL smoothly at
$x_*(1)\approx 4\times10^{-8}$ and $x_*(2)=0.97$, so there should be
another regime of $0.97<x<1=x_2$ inside condition II. The first four
right cross points of type2-type1 solutions are $1.65$, $0.743$,
$1.1$ and $0.97$. By inference, the cross points of the fifth and
following type2-type1 solutions will be narrowly located around
$x=1$.
When the infinite number of type2-type1 solutions are taken into
account, there will be fine structures around $x=1$ in subsection
3.3.1 and Figure \ref{fig:typealpha_x_*_x_0}. However, the solution
behaviours of crossing the SCL smoothly under condition of fine
structure are like those under conditions II, III and IV near $x=1$.

\subsection[]{Type $\mathcal{Z}$ Voids:
ISSV Solutions with Shocks}

Shock phenomena are common in various astrophysical flows, such as
planetary nebulae, supernova remnants, and even galaxy clusters gas
(e.g. Castor et al. 1975; McNamara et al. 2005). In this subsection,
we present type $\mathcal{Z}$ ISSV solutions, namely, self-similar
void solutions with shocks. Equations (\ref{shockcm}) and
(\ref{shockcmom}) are mass and momentum conservations. The
isothermality is a strong energy requirement. In our isothermal
model, an example of polytropic process, the energy process is
simplified. This simplification gives qualitative or
semi-quantitative description of the energy process for a shock
wave. By introducing parameter $\tau$ for the temperature difference
after and before a shock, we can describe more classes of shocks
(Bian \& Lou 2005).

The basic procedure to construct a spherical ISSV solution with
shocks is as follows. Given $(x_0,\ v_0=x_0,\ \alpha_0)$ at the void
boundary, we can integrate ODEs (\ref{ODE1}) and (\ref{ODE2})
outwards from $x_0$; in general, numerical solutions cannot pass
through the SCL smoothly (if they do, they will be referred to as
type $\mathcal{X}$ ISSV solutions); however, with an outgoing shock,
solutions can readily cross the SCL (e.g. Tsai \& Hsu 1995; Shu et
al. 2002; Shen \& Lou 2004; Bian \& Lou 2005); finally global
($x_0<x<+\infty$) solutions can be constructed by a combination of
integration from $x_0$ to $x_{ds}$, shock jump and integration from
$x_{us}$ to $+\infty$, where $x_{ds}$ and $x_{us}$ are defined in
subsection 2.2 as the radial expanding velocity of a shock on the
downstream and upstream sides, respectively.

A typical ISSV solution with a shock has four degrees of freedom
(DOF) within a sensible parameter regime. For example, we need
independent input of $x_0$, $\alpha_0$, $x_{ds}$ and $\tau$
to determine an ISSV solution with a shock, while the degree of
freedom for type $\mathcal{X}$ solutions is one (i.e. $x_*$ plus the
type of eigen-derivative crossing the SCL are enough to make a type
$\mathcal{X}$ ISSV solution). When we consider the simplest
condition that $\tau=1$,
the DOF of a void solution with shocks is three. So infinite void
shock solutions exist. By fixing one or two parameters, we can
enumerate all possible values of the other parameter to obtain all
possible ISSV solutions. For example, by fixing velocity parameter
$V$,
we can adjust mass parameter $A$ and $x_{us}$ to explore all void
solutions (see equation \ref{largeasymp}); following this
procedure, $x_0$, $\alpha_0$ and $x_{ds}$ are determined by $V$,
$A$ and $x_{us}$. There is a considerable freedom to set up an
ISSV solution with shocks.
In astrophysical flows, we would like to learn the expansion speed
of a void, the density surrounding a void and the radial speed of
gas shell at large radii. We then choose one or two parameters in
$x_0$, $\alpha_0$ and $V$ as given parameters to search for ISSV
solutions by changing other parameters such as $x_{us}$.

In Section 3.4.1, we will first consider the simple case:
equi-temperature shock void (i.e. $\tau=1$), and refer to as void
solutions with equi-temperature
shocks or type
$\mathcal{Z}_{\rm I}$ ISSV solutions. Several type $\mathcal{Z}_{\rm
I}$ voids with different behaviours near void boundaries and outer
envelopes (a static envelope, outflows and inflows) will be
presented. Phase diagram matching method will be described and
extended to the so-called `phase net', with the visual convenience
to search for ISSV solutions with more DOF. Section 3.4.2 presents
type $\mathcal{Z}_{\rm II}$ ISSV solutions: void solutions with
two-soundspeed shocks (i.e. $\tau>1$).

\subsubsection[]{Type $\mathcal{Z}_{\rm I}$ Void Solutions: Voids\\
\qquad\ \ with Same-Soundspeed Shocks}

For an equi-temperature
shock, we have $\tau=1$ and
thus $x_{ds}=x_{us}=x_s$ (see Section 2.2). As already noted, the
DOF of $\mathcal{Z}_{\rm I}$ void solutions is three. We can use
$\{x_0,\ \alpha_0$, $\ x_s\}$
to construct type $\mathcal{Z}_{\rm I}$ ISSV solutions.

We have freedom to set the condition $(x_0,\ \alpha_0)$ at the void
edge to integrate outwards. Before reaching the SCL, we set an
equi-temperature
shock at a fairly arbitrary $x_s$ to
cross the SCL. We then combine the integrations from $x_0$ to
$x_s^-$ and from $x_s^+$ to $+\infty$ by a shock jump to form a type
$\mathcal{Z}_{\rm I}$ ISSV solution with a shock.

We emphasize that under type $\mathcal{Z}_{\rm I}$ condition, the
insertion of an equi-temperature
shock does assure that void solutions jump across the SCL. Physics
requires that at every point $(x,\ v(x),\ \alpha(x))$ of any void
solution, there must be $x>v(x)$ for a positive mass. Equation
(\ref{shock1tau2}) indicates that across the equi-temperature
shock front, the product of two negative $v_{d}-x_s$ and $v_{u}-x_s$
makes 1. In our model, $x_s-v_d<1$ so $x_s-v_u$ must be larger than
$1$. So the downstream and upstream are separated by the SCL.
However, in type $\mathcal{Z}_{\rm II}$ condition with $\tau>1$,
this special property may not always hold on. We shall require such
a separation across the SCL as a necessary physical condition.

Figure \ref{fig:type_beta_I} shows several type $\mathcal{Z}_{\rm
I}$ ISSV shock solutions with void expansion
at half sound speed and $0.03$ times sound
speed. From this figure, we know that even the voids expand at the
same speed, with different density $\alpha_0$ near the void edge and
the radial velocities of the shock wave ($x_s$), they can have outer
envelopes of various dynamic behaviours. From values of $v(x)$ and
$\alpha(x)$ at large $x$ (say 10), we can estimate $V$ and $A$.
Different from type $\mathcal{X}$ ISSV solutions, some type
$\mathcal{Z}_{\rm I}$ ISSV shock solutions have outer envelopes with
a negative $V$ (e.g. curves $1'$ and $2'$ in Fig.
\ref{fig:type_beta_I}), that is, contracting outer envelopes.
Numerical calculations show that voids with contracting envelopes
usually expand very slowly (voids $1'$ and $2'$ in Fig.
\ref{fig:type_beta_I} expand at $0.03a$).

\begin{figure}
\centering
\includegraphics[width=0.50\textwidth]{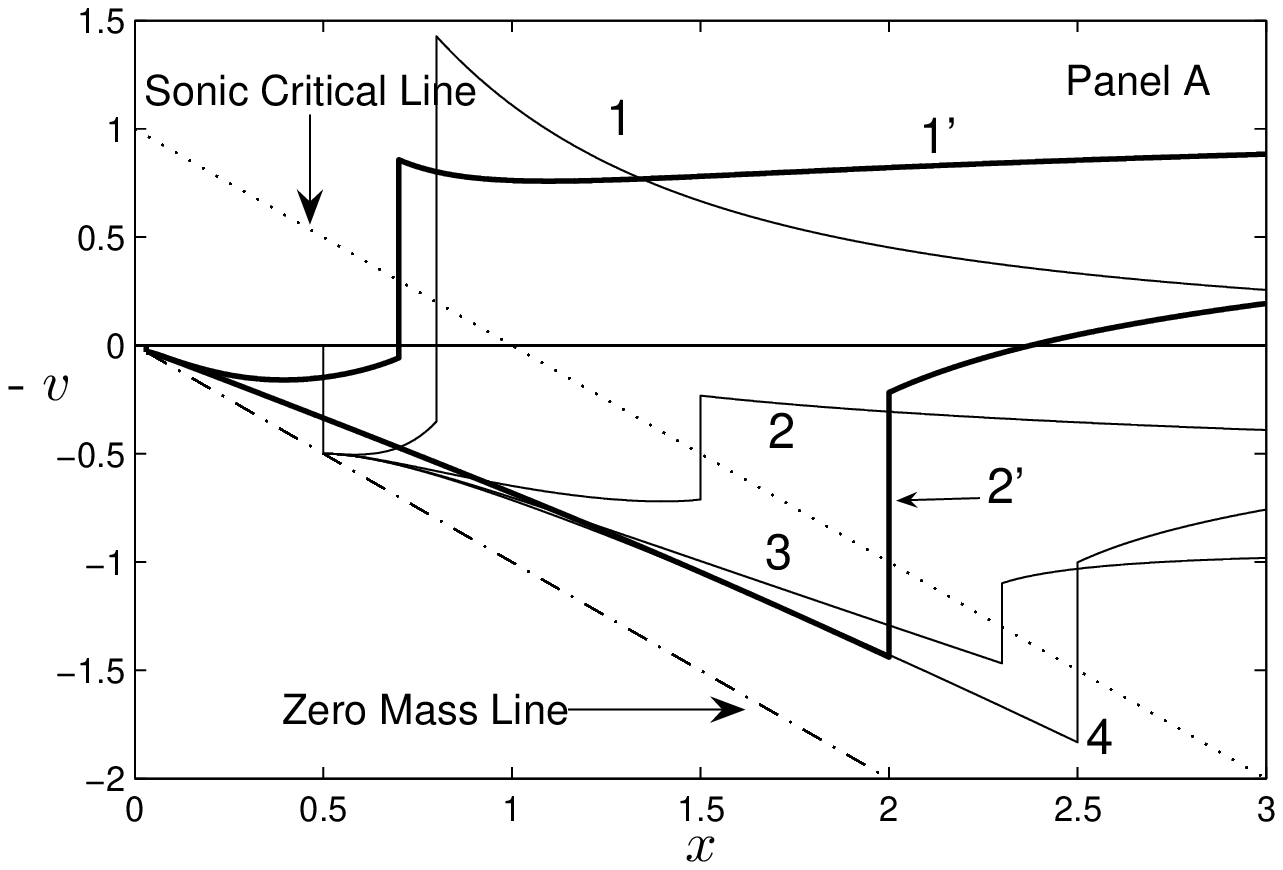}
\includegraphics[width=0.50\textwidth]{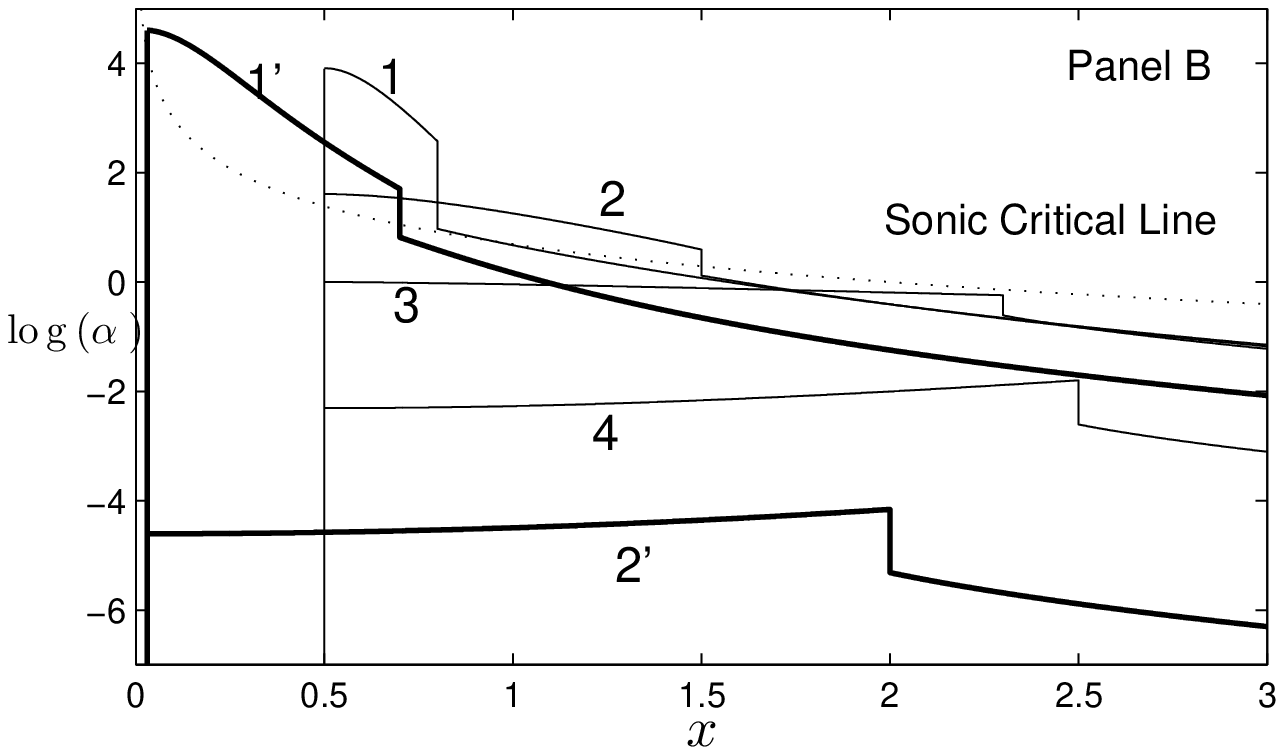}
\caption{We show six type $\mathcal{Z}_{\rm I}$ void shock
solutions. Four of them have same $x_0=0.5$ but different values
of $\alpha_0$ marked by numbers $1, 2, 3, 4$. The other two have
same $x_0=0.03$ but different values of $\alpha_0$ marked by
number $1', 2', 3', 4'$. Panel A above presents $-v(x)$ versus $x$
profiles and panel B below shows $\log\alpha$ versus $x$ profiles.
Values of the two free parameters $(\alpha_0,\ x_s)$ are
$(50,\ 0.8)$,
$(5,\ 1.5)$,
$(1,\ 2.3)$,
$(0.1,\ 2.5)$ for curves $1,2,3,4$, respectively; and
$(100,\ 0.7)$,
$(0.01,\ 2)$ for curves $1'$ and $2'$, respectively. The dotted
curves in both panels are Sonic Critical
Line.}\label{fig:type_beta_I}
\end{figure}

\begin{figure}
\centering
\includegraphics[width=0.50\textwidth]{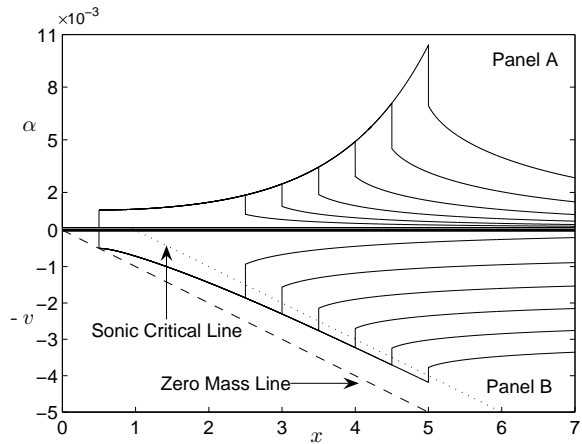}
\caption{Several type $\mathcal{Z}_I$ ISSV solutions with
quasi-smooth edges in light solid curves with the same values of
$x_0=0.5$ and $\alpha_0=10^{-3}$ but different shock speed $x_s$
from $2.5$ to $5.0$ every $0.5$ in step. Panel A presents
$\alpha(x)$ versus $x$ profiles and panel B shows $-v(x)$ versus
$x$ profiles. The dotted and dash lines in panel B are the SCL and
the ZML, respectively.}\label{fig:type_beta_I_quasismooth}
\end{figure}

In addition to shocks and outer envelope dynamics, another important
difference between type $\mathcal{Z}_I$ and type $\mathcal{X}$ ISSV
solutions lies in behaviours of shells surrounding the void edge.
From Section 3.3 (see Fig. \ref{fig:typealphaI}),
we have known that the reduced mass density of type $\mathcal{X}$
void solutions must encounter a sharp jump and decrease
monotonically with increasing $x$. However, Fig.
\ref{fig:type_beta_I} (curves $4$ and $2'$ with the y-axis in
logarithmic scales) and Fig. \ref{fig:type_beta_I_quasismooth}
indicate that with shocks involved, the density of shells near void
edges can increase with increasing $x$. Under these conditions,
density jumps from a void to gas materials around voids appear not
to be very sharp. Voids described by solutions like curves in Figure
\ref{fig:type_beta_I_quasismooth} and curves $4$ and $2'$ in Figure
\ref{fig:type_beta_I} have such `quasi-smooth' edges. These
solutions can approximately describe a void with a smooth edge,
whose outer shell gradually changes from vacuum in the void to gas
materials, without sharp density jump. Or, they describe a void with
a quasi-smooth edge, whose outer shell gradually changes from vacuum
in the void to gas materials, with a small density jump.
Figure \ref{fig:type_beta_I_quasismooth} clearly shows that the
faster a shock moves relative to void edge expansion, the higher
density rises from void edge to shock.



\paragraph{Type $\mathcal{Z}_{\rm I}$ Voids with a SIS Envelope}

Void phenomena in astrophysics indicate an expanding void in the
centre and static gas medium around it in the outer space. For
example, a supernova explodes and ejects almost all its matter into
space.
If the shock explosion approximately starts from the central core of
the progenitor star, the remnant of the supernova is then
approximately spherically symmetric and a void may be generated
around the explosion centre (e.g. Lou \& Cao 2008). If the gravity
of the central compact object may be ignored, we then describe this
phenomenon as an expanding spherical void surrounded by a static
outer envelope. The analysis of Section 3.3.3 indicates that all
type $\mathcal{X}$ ISSV solutions cannot describe this kind of
phenomena. However with rebound shocks, it is possible to construct
a model for an expanding void surrounded by a static SIS envelope.

Shu (1977) constructed the expansion-wave collapse solution (EWCS)
to describe a static spherical gas with an expanding region
collapsing towards the centre. In fact, EWCS outer envelope with
$x>1$ is the outer part of a SIS solution (see equation \ref{SIS}).
We now construct several ISSV solutions with an outer SIS envelope.
An outer SIS envelope has fixed two DOF of a type $\mathcal{Z}_{\rm
I}$ ISSV solution, that is, $V=0$ and $A=2$, so there is only one
DOF left. A simple method is to introduce a shock at a chosen point
$x_s$ of EWCS solution except $(x=1,\ v=0,\ \alpha=1)$ (we emphasize
that only one point $(x=1,\ v=0,\ \alpha=1)$ of the EWCS solution is
at the SCL and all the other points lay on the upper right to the
SCL in the $-v(x)$ versus $x$ profile) and make the right part of
EWCS solution the upstream of a shock, then we can obtain $(v_{ds},\
\alpha_{ds})$ on the downstream side of a shock. If the integration
from $(x_s,\ v_{ds},\ \alpha_{ds})$ leftward touches the ZML at
$x_0$, a type $\mathcal{Z}_{\rm I}$ ISSV solution with a static
outer envelope is then constructed.

We introduce the $\alpha-v$ phase diagram to deal with the
relationship among the free parameters and search for
eigensolutions of ODEs (\ref{ODE1}) and (\ref{ODE2}). Hunter
(1977) introduced this method to search for complete self-similar
eigensolutions of two ODEs. Whitworth \& Summer (1985) used this
method to combine free-fall solutions and LP solutions in the
centre with certain asymptotic solutions at large radii. Lou \&
Shen (2004) applies this method to search for eigensolutions of
ODEs (\ref{ODE1}) and (\ref{ODE2}) which can cross the SCL twice
smoothly. In the case of Lou \& Shen (2004), the DOF is 0. So
there is an infinite number of discrete eigensolutions.

\begin{figure}
\centering
\includegraphics[width=0.5\textwidth]{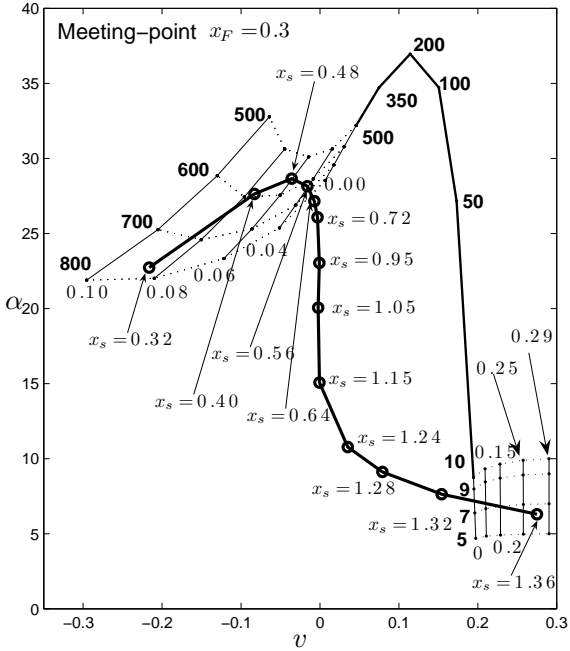}
\caption{The phase diagram of $\alpha$ versus $v$ at a chosen
meeting point $x_F=0.3$ for match of the SIS solution (e.g., Shu
1977) and type $\mathcal{Z}$ void shock solutions. Each open circle
symbol joint by a heavy solid curve denotes an integration from a
chosen $x_s$ (marked besides each open circle) towards $x_F<x_s$. We
impose the equi-temperature
shock conditions at $x_s$
from $0.32$ to $1.36$ for the outer SIS solution and then integrate
from $x_s$ back to $x_F$ to get $[v(x_F),\ \alpha(x_F)]$ as marked
in the phase diagram. The change of $x_s$ naturally leads to a phase
curve shown here
by the heavy solid curve. We choose the range of $x_s$ to be
$[0.32,\ 1.36]$ because $x_s$ must be larger than $x_F$ and a larger
$x_s$ than $1.36$ will give rise to solution curves encountering the
SCL at $x\geq x_F=0.3$.
Meanwhile, two ``phase nets", made of the light solid curves and
dotted curves and connected by a medium heavy solid curve, are
actually generated by integrations from chosen $(x_0,\ v_0=x_0,\
\alpha_0)$. Each solid curve in the two nets (including the
connecting medium heavy solid curve) is an equal$-x_0$ curve, that
is, every point in the same solid curve corresponds to the same
value of $x_0$ noted besides each equal$-x_0$ curve; and each dotted
curve in the two nets is an equal$-\alpha_0$ curve with the value of
$\alpha_0$ in boldface noted besides each curve.
For the lower right net, the points in the net correspond to an
initial condition of $(x_0,\
\alpha_0)\in\{x_0|x_0\in[0,0.29]\}\times\{\alpha_0|\alpha_0\in[5,10]\}$;
the value of $x_0$ is $0$, $0.15$, $0.2$, $0.25$, $0.29$
respectively from the left equal-$x_0$ solid curve to the right one
and the value of $\alpha_0$ is $5$, $7$, $9$, $10$ respectively from
the bottom equal-$\alpha_0$ dotted curve to the top one. For the
upper left net, the points in the net correspond to an initial
condition of $(x_0,\
\alpha_0)\in\{x_0|x_0\in[0,0.1]\}\times\{\alpha_0|\alpha_0\in[500,800]\}$;
the value of $x_0$ is $0$, $0.04$, $0.06$, $0.08$, $0.10$
respectively from the right equal-$x_0$ solid curve to the left one
and the value of $\alpha_0$ is $500$, $600$, $700$, $800$
respectively from the top equal-$\alpha_0$ dotted curve to the
bottom one. The medium solid curve is an equal-$x_0$, too, with
$x_0=0$ and $\alpha_0=10$, $50$, $100$, $200$, $350$ and $500$ from
lower right to left. This medium solid curve shows the trend of
phase curves as we increase $\alpha_0$ value in large steps.
}\label{fig:phasediagram_void_SIS}
\end{figure}

For type $\mathcal{Z}_{\rm I}$ ISSV shock solutions with a static
SIS outer envelope, the DOF is one. We insert a shock at $x_s$ in
the SIS and then integrate inwards from $x_s$ to a fixed meeting
point $x_F$. Adjusting the value of $x_s$ will lead to a phase curve
$[v(x_F)^+,\ \alpha(x_F)^+]$ in $\alpha$ versus $v$ phase diagram.
Meanwhile, an outward integration from a chosen void boundary
condition $(x_0,\ v_0=x_0,\ \alpha_0)$ reaches a phase point
$[v(x_F)^-,\ \alpha(x_F)^-]$ at $x_F$. Varying values of $x_0$ or
$\alpha_0$ will lead to another phase curve $[v(x_F)^-,\
\alpha(x_F)^-]$ in the $\alpha$ versus $v$ phase diagram. We note
that changing both values of $x_0$ and $\alpha_0$ will result in a
``phase net" (i.e. a two-dimensional mesh of phase curves) in
$\alpha$ versus $v$ diagram (see Fig.
\ref{fig:phasediagram_void_SIS}). If such a ``phase net" of $(x_0,\
\alpha_0)$ and the phase curve of $x_s$ share common points
(usually, there will be an infinite number of common points
continuously as such ``phase net" is two dimensional), type
$\mathcal{Z}_{\rm I}$ ISSV shock solutions with a static SIS
envelope can be constructed.

Figure \ref{fig:phasediagram_void_SIS} presents the phase diagram at
a meeting point $x_F=0.3$ to search for type $\mathcal{Z}_{\rm I}$
ISSV shock solutions with an outer SIS envelope. Note that part of
the phase curve falls into the phase net, revealing that an infinite
number of type $\mathcal{Z}_{\rm I}$ ISSV shock solutions with outer
SIS envelope indeed exist continuously. Shown by Figure
\ref{fig:phasediagram_void_SIS}, numerical results suggest that when
shock position $x_s>0$ is less than $0.62$ or larger than $1.335$,
there is at least one type $\mathcal{Z}_{\rm I}$ ISSV that can exist
in the downstream side of a shock. However, if a shock expands at a
radial velocity between $0.62a$ and $1.335a$, it is impossible for a
type $\mathcal{Z}_{\rm I}$ ISSV to exist inside a shock with a SIS
envelope. Table 4 contains values of $x_0$, $\alpha_0$ and $x_s$ of
some typical type $\mathcal{Z}_{\rm I}$ ISSV shock solutions with a
SIS envelope. Figure \ref{fig:phase_void_SIS_x_0_alpha_0} is a phase
diagram showing how $x_0$ and $\alpha_0$ are evaluated with $x_s$
changing to construct a type $\mathcal{Z}_{\rm I}$ ISSV shock
solutions with an outer SIS envelope.

\begin{table}
\tabcolsep 0pt \caption{Values of $x_0$, $\alpha_0$ and $x_s$ for
several type $\mathcal{Z}_{\rm I}$ void shock solutions with an
outer static SIS envelope are summarized here}\vspace*{-12pt}
\begin{center}
\def\temptablewidth{0.3\textwidth}
{\rule{\temptablewidth}{1pt}}
\begin{tabular*}{\temptablewidth}{@{\extracolsep{\fill}}ccc}
$x_0   $ & $\alpha_0     $ & $x_s$ \\    \hline
$0.018 $ & $4.3\times10^6$ &$0.02$\\
$0.063 $ & $1.5\times10^4$ &$0.1$\\
$ 0.09 $ & $1.1\times10^3$ &$0.26$\\
$0.077 $ & $5.9\times10^2$ &$0.4$\\
$ 0.01 $ & $9.5\times10^2$ &$0.62$\\
$<0.005$ & $7.9$           &$1.335$\\
$ 0.12 $ & $7.5$           &$1.338$\\
$ 0.27 $ & $6.3$           &$1.36$\\
$ 0.65 $ & $4.1$           & $1.5$\\
$ 2.65 $ & $2.3$           &$3.00$
       \end{tabular*}
       {\rule{\temptablewidth}{1pt}}
       \end{center}
       \end{table}

\begin{figure}
\centering
\includegraphics[height=6cm,width=0.5\textwidth]{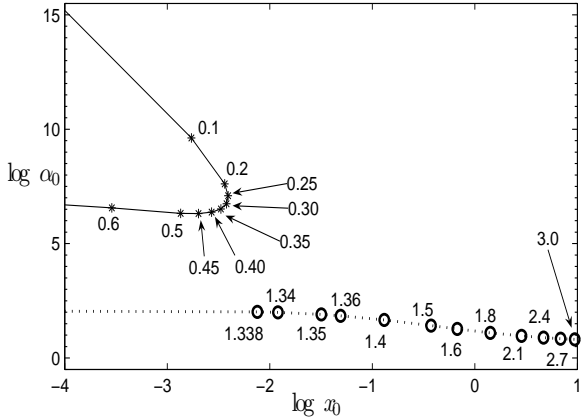}
\caption{The phase diagram of $\log\alpha_0$ versus $\log x_0$
shows the relationship among $x_s$, $x_0$ and $\alpha_0$ of type
$\mathcal{Z}_{\rm I}$ void shock solutions with a static SIS
envelope. 
The DOF of these three parameters is only one; i.e. given
arbitrary one of the three, the other two parameters are
determined in constructing a type $\mathcal{Z}_{\rm I}$ void shock
solutions with a static SIS envelope. For any point in the two
curves, $x_0$ and $\alpha_0$ indicated by its $x$ and $y$
coordinates with $x_s$ marked, correspond to a type
$\mathcal{Z}_{\rm I}$ void shock solution. The upper left solid
curve with its data points in asterisk symbol corresponds to the
condition $x_s<0.62$ and the lower dotted curve with its data
points in open circle corresponds to the condition $x_s>1.335$.
The first condition referred to as Class I gives its largest
$x_0=0.09$ when $x_s=0.26$. Although $x_0$ for the second
condition referred to as Class II ranges along the entire real
axis, it usually takes a relative large value; moreover, in Class
II solutions, the reduced density $\alpha_0$ at the void boundary
is in the order of unity. }\label{fig:phase_void_SIS_x_0_alpha_0}
\end{figure}

All of Fig. \ref{fig:phasediagram_void_SIS}, Fig.
\ref{fig:phase_void_SIS_x_0_alpha_0} and Table 4 clearly indicate
type $\mathcal{Z}_{\rm I}$ ISSV shock solutions with a SIS envelope
can be generally divided into two classes according to $x_s$. Class
I type $\mathcal{Z}_{\rm I}$ void shock solutions with an outer
static SIS envelope have $x_s<0.62$ usually with a smaller value of
$x_0$ and a higher value of $\alpha_0$. Class II solutions have
$x_s>1.335$ usually with a larger value of $x_0$ and a medium value
of $\alpha_0$. By a numerical exploration, the maximum of $x_0$ of
Class I solutions is $\sim 0.09$ for $x_s=0.26$ and
$\alpha_0=1.1\times10^3$; these voids expand at a low speed of $<
0.1a$; and the reduced density $\alpha_0$ at the void boundary is
usually $>10^2$, indicating a sharp edge density peak. Finally, we
note that Class I voids involve shocks expanding at subsonic speeds.
In this situation, the outer region $x>x_s$ is not
completely static. The SIS envelope only exists at $x\geq 1$, and
the region between $x_s$ and $1$
is a collapse region (see two class I ISSV shock solutions 1 and 2
in Fig. \ref{fig:void_SIS}). While in Class II ISSV shock solutions,
shock expands supersonically and with $x_0$
usually relatively large.
So the upstream side of a shock in Class II ISSV solutions is static
(see two class II ISSV solutions 3 and 4 shown in Fig.
\ref{fig:void_SIS}).
In rare situations, $x_0$ can be small (e.g. $x_s<1.4$) and
$\alpha_0$ neither large nor small, indicating that Class II voids
have moderately sharp edges.

\begin{figure}
\centering
\includegraphics[height=12cm,width=0.5\textwidth]{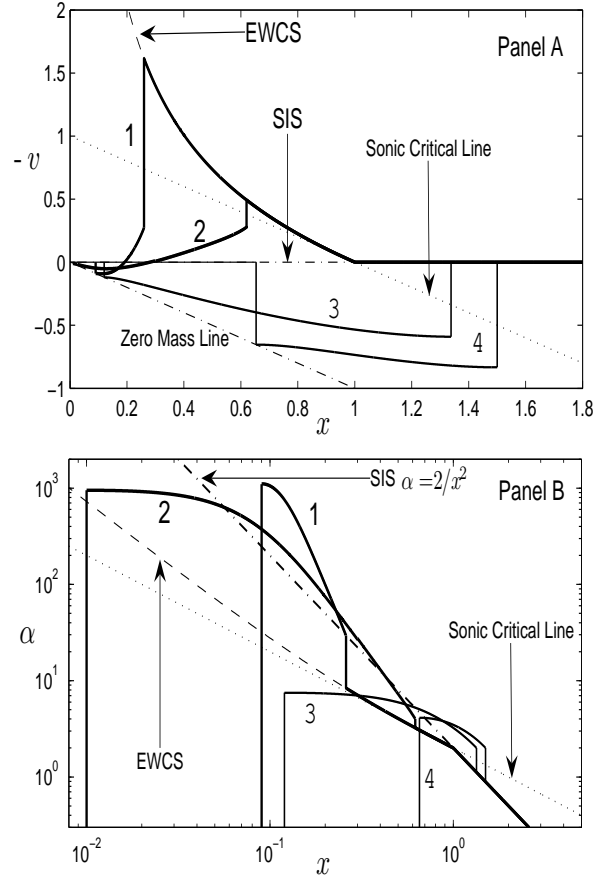}
\caption{Four typical type $\mathcal{Z}_{\rm I}$ void shock
solutions with static SIS envelope. The two heavy solid curves in
both panels are the Class I type $\mathcal{Z}_{\rm I}$ void
solutions with a SIS envelope and a subsonic shock (i.e.,
$x_s<1$), and the two light solid curves in both panels are the
Class II type $\mathcal{Z}_{\rm I}$ void solutions with an outer
SIS envelope and a supersonic shock (i.e., $x_s>1$). Panel A
presents $-v(x)$ versus $x$ profiles. Panel B presents $\alpha$
versus $x$ profiles using a logarithmic scale along the $y-$axis.
The key data $(x_0,\ \alpha_0,\ x_s)$ of these solutions are
$(0.09,\ 1.1\times10^3,\ 0.26)$,
$(0.01,\ 9.5\times10^2,\ 0.62)$,
$(0.12,\ 7.5,\ 1.338)$, and
$(0.65,\ 4.1,\ 1.50)$ for curves 1, 2, 3, and 4, respectively. The
dash curves in both panels are part of the EWCS solution.
}\label{fig:void_SIS}
\end{figure}

\paragraph{Type $\mathcal{Z}_{\rm I}$ voids with expanding envelopes:
breezes, winds, and outflows}

In our ISSV model, we use parameters $V$ and $A$ to characterize
dynamic behaviours of envelopes. Equation (\ref{largeasymp})
indicates that $V>0$ describes an expanding envelope at a finite
velocity of $Va$, and a larger $A$ corresponds to a denser
envelope. For $V=0$, the expansion velocity vanishes at large
radii, corresponding to a breeze; a smaller $A$ than $2$ is
required to make sure an outer envelope in breeze expansion. For
$V>0$, the outer envelope is a wind with finite velocity at large
radii.

We apply the similar method to construct type $\mathcal{Z}_{\rm I}$
voids with expanding envelopes as we deal with type
$\mathcal{Z}_{\rm I}$ voids with outer SIS envelopes. The difference
between the two cases is that $V$ and $A$ are allowed to be
different from $V=0$ and $A=2$.
In this subsection, we usually choose a meeting point $x_F$ between
$x_0$ and $x_s$. Then by varying $x_0$ and $\alpha_0$,
we obtain a phase net composed by $[v(x_F)^-,\ \alpha(x_F)^-]$.
Given $V$, we adopt $A$ and integrate ODEs (\ref{ODE1}) and
(\ref{ODE2}) from large $x$
towards $x_s$. After setting a shock at $x_s$, we integrate ODEs
towards $x_F$. By varying $A$ and $x_s$, we obtain a phase net
composed by $[v(x_F)^+,\ \alpha(x_F)^+]$. The overlapped area of two
phase nets reveals the existence of type $\mathcal{Z}_{\rm I}$ ISSV
with dynamic envelopes characterized by $V$ and $A$.


\begin{figure}
\centering
\includegraphics[width=0.5\textwidth]{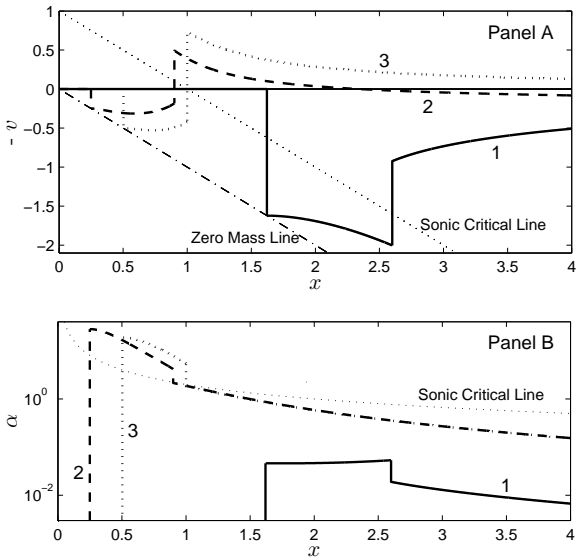}
\caption{Three typical type $\mathcal{Z}_{\rm I}$ void shock
solutions with different outer envelopes. Panel A presents $-v(x)$
versus $x$ profiles. Panel B presents $\alpha(x)$ versus $x$
profiles in a logarithmic scale along the $y-$axis. The heavy solid
curve $1$ gives a type $\mathcal{Z}_{\rm I}$ void shock solution
with a quite thin breeze outer envelope whose
$(x_0,\ \alpha_0,\ x_s,\ V,\ A)=(1.62,\ 0.046,\ 2.6,\ 0,\ 0.1)$.
The heavy dash curve $2$ gives a type $\mathcal{Z}_{\rm I}$ void
shock solution with a wind outer envelope whose
$(x_0,\ \alpha_0,\ x_s,\ V, \ A)=(0.25,\ 28.2,\ 0.90,\ 0.2,\ 2.5)$.
The heavy dotted curve $3$ gives a type $\mathcal{Z}_{\rm I}$ void
shock solution with an accretion outer envelope whose
$(x_0,\ \alpha_0,\ x_s,\ V, \ A)=(0.50,\ 18.8,\ 1.00,\ 0,\ 2.50)$.
The monotonic dotted curves in both panels stand for the
SCL.}\label{fig:type_beta_I_envelop}
\end{figure}

Figure \ref{fig:type_beta_I_envelop} gives two examples of type
$\mathcal{Z}_{\rm I}$ voids with breeze and wind. We emphasize
that $A>2$ is required in such ISSV solutions with an outer
envelope wind and subsonic shock. Actually, the larger the
velocity parameter $V$ is, the smaller the mass parameter $A$ is
needed. Large $A$ is required to guarantee that the upstream
region of a global solution is on the upper right part of the SCL
when the shock moves subsonically (i.e. $x_s<1$). Physically, if
an ISSV is surrounded by a subsonic shock wave, the wind outside
needs to be dense enough.

\paragraph{Type $\mathcal{Z}_{\rm I}$ voids with
contracting outer envelopes: accretions and inflows}

We explore ISSV with contracting envelopes, such as accretion
envelopes.
Type $\mathcal{X}_{\rm I}$ voids under conditions I and II have
contracting envelopes. These type $\mathcal{X}_{\rm I}$ voids are
all surrounded by very dense shells with density decreasing with
increasing radius. With shocks involved, central voids can have
envelopes of various properties. Type $\mathcal{Z}_{\rm I}$ voids
with contracting outer envelopes are also studied. To have a
contracting envelope, the velocity parameter $V$ should be negative
or approach $0^{-}$. A negative $V$ and a positive $A$ describe an
outer envelope inflowing at a velocity of $aV$ from large radii. For
$V=0$ and $A>2$, an outer envelope has an inflow velocity vanishing
at large radii (Lou \& Shen 2004; Bian \& Lou 2005).
Figure \ref{fig:type_beta_I_envelop} gives examples of type
$\mathcal{Z}_{\rm I}$ voids with accreting outer envelopes.

\subsubsection[]{Type $\mathcal{Z}_{\rm II}$ ISSV Solutions:
Voids Surrounded by Two-Soundspeed Shocks in Envelopes}

In the previous section, we explored type $\mathcal{Z}_{\rm I}$ ISSV
solutions featuring the equi-temperature
shock.
There, $\tau=1$ indicates the same sound speed $a$ across a shock.
The isothermal sound speed $a$ can be expressed as
\begin{equation}\label{soundspeed}
a=\left(\frac{p}{\rho}\right)^{1/2}
=\left[\frac{(Z+1)k_B}{\mu}T\right]^{1/2}\ ,
\end{equation}
where $k_B$ is Boltzmann's constant,
$\mu$ is the mean atomic mass} and $Z$ is the ionization state.
$Z=0$ corresponds to a neutral gas and $Z=1$ corresponds to a fully
ionized gas.
In various astrophysical processes, shock waves increase the
downstream temperature, or change the proportion of gas particles;
moreover, the ionization state may change after a shock passage
(e.g. champagne flows in H$_{\rm II}$ regions, Tsai \& Hsu 1995; Shu
et al. 2002; Bian \& Lou 2005; Hu \& Lou 2008). Such processes
lead to two-soundspeed shock waves with $\tau>1$. In this section,
we consider $\tau>1$ for type $\mathcal{Z}_{\rm II}$ ISSV shock
solutions with two-soundspeed shocks. Global $\mathcal{Z}_{\rm II}$
void solutions have
temperature changes across shock waves while both downstream and
upstream sides remain isothermal, separately. With a range of
$\tau>1$, it is possible to fit our model to various astrophysical
flows.

\begin{figure} \centering
\includegraphics[height=10cm,width=0.5\textwidth]{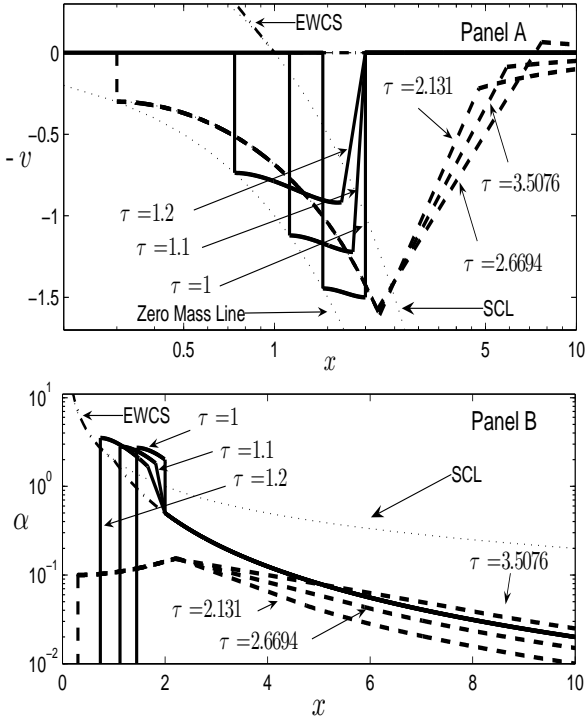}
\caption{Six type $\mathcal{Z}$ ISSV solutions with shocks. Panel A
presents $-v(x)$ versus $x$ profiles and panel B presents
$\alpha(x)$ versus $x$ profiles. The heavy solid curves labelled by
their corresponding $\tau$ in both panels are type $\mathcal{Z}$
ISSV shock solutions with $x_s=2$ and $\tau=1,\ 1.1,\ 1.2$
respectively. The $\tau=1$ case is a type $\mathcal{Z}_{\rm I}$ ISSV
shock solution and the latter two with $\tau>1$ are type
$\mathcal{Z}_{\rm II}$ ISSV shock solutions. The heavy dash curves
labelled by their corresponding $\tau$ in both panels are solutions
with reduced velocity $V=0$ and $A=1,\ 1.5,\ 2.5$ respectively. The
one with $A=2.5$ has an envelope accretion and the other two have
envelope breezes. The relevant data of these six solutions are
summarized in Table 5. Shock jumps of type $\mathcal{Z}_{\rm II}$
ISSV solutions do not appear vertical as those of type
$\mathcal{Z}_{\rm I}$ ISSV shock solutions (e.g. Bian \& Lou 2005)
because of different sound speeds across a shock front and thus
different scales of reduced radial $x\equiv
r/(at)$.}\label{fig:betaii}
\end{figure}

For $\tau>1$, the DOF of type $\mathcal{Z}_{\rm II}$ ISSV shock
solutions is four, i.e. one more than that of type $\mathcal{Z}_{\rm
I}$ ISSV shock solutions with $\tau=1$.
We will not present details to construct type $\mathcal{Z}_{\rm
II}$ ISSV shock solutions as they differ from the corresponding
type $\mathcal{Z}_{\rm I}$ ISSV shock solution only in the
quantitative sense. General properties such as the behaviours near
the void boundary and the outer envelope of type $\mathcal{Z}_{\rm
II}$ ISSV shock solutions remain similar to those of type
$\mathcal{Z}_{\rm I}$ ISSV shock solutions. We present examples of
typical type $\mathcal{Z}_{\rm II}$ ISSV shock solutions in Figure
\ref{fig:betaii}.

\begin{table}
\tabcolsep 0pt \caption{Parameters of several type $\mathcal{Z}_{\rm
I}$ and $\mathcal{Z}_{\rm II}$ ISSV shock solutions shown in Figure
\ref{fig:betaii}}\vspace*{-12pt}
\begin{center}
\def\temptablewidth{0.45\textwidth}
{\rule{\temptablewidth}{1pt}}
\begin{tabular*}{\temptablewidth}{@{\extracolsep{\fill}}ccccccc}
$x_0$     &  $\alpha_0$        & $x_{ds}$ &$x_{us}$&$\tau$& $V$ &  $A$ \\
\hline
$1.45$    &  $2.72$            & $2$      & $2$   & $1$   & $0$ &  $2$\\
$1.12$    &  $2.81$            & $1.82$   & $2$   & $1.1$ & $0$ &  $2$\\
$0.74$    &  $3.51$            & $1.67$   & $2$   & $1.2$ & $0$ &  $2$\\
$2.65$    &  $2.26$            & $3$      & $3$   & $1$   & $0$ &  $2$\\
$1.21$    &  $1.87$            & $2$      & $3$   & $1.5$ & $0$ &  $2$\\
$0.3$     &  $0.1$             & $2.2303$ & $4.7528$&$2.131$&$0$&  $1$  \\
$0.3$     &  $0.1$             & $2.2064$ & $5.8898$&$2.6694$&$0$&  $1.5$\\
$0.3$     &  $0.1$             & $2.1886$ & $7.6767$&$3.5076$&$0$&  $2.5$\\
       \end{tabular*}
       {\rule{\temptablewidth}{1pt}}
       \end{center}
       \end{table}


\section[]{Astrophysical Applications}

\subsection[]{The role of self-gravity in gas clouds}

Earlier papers attempted to build models for hot bubbles and
planetary nebulae (e.g. Weaver et al. 1977; Chevalier 1997a) without
including the gas self-gravity. In reference to Chevalier (1997a)
and without gravity, our nonlinear ODEs (\ref{ODE1}) and
(\ref{ODE2}) would then become
\begin{equation}\label{ODE1_noG}
\left[(x-v)^2-1\right]\frac{dv}{dx}=-\frac{2}{x}(x-v)\ ,
\end{equation}
\begin{equation}\label{ODE2_noG}
\left[(x-v)^2-1\right]\frac{1}{\alpha}\frac{d\alpha}{dx}=-\frac{2}{x}(x-v)^2\
.
\end{equation}
ODEs (\ref{ODE1_noG}) and (\ref{ODE2_noG}) allow both outgoing and
inflowing outer envelopes around expanding voids.\footnote{Actually,
Chevalier (1997a) did not consider physically possible situations of
contracting outer envelopes.}
Self-similar solutions of ODEs (\ref{ODE1_noG}) and
(\ref{ODE2_noG}) cannot be matched via shocks with a static solution
of uniform mass density $\rho$.
For comparison, the inclusion of
self-gravity can lead to a static SIS.
In some
circumstances, there may be no apparent problem when ODEs
(\ref{ODE1_noG}) and (\ref{ODE2_noG}) are applied to describe
planetary nebulae because AGB wind outer envelope may have finite
velocity at large radii. However, in the interstellar bubbles
condition, a static ISM should exist outside the interaction region
between stellar wind and ISM. In Weaver et al. (1977), a static
uniformly distributed ISM surrounding the central bubble was
specifically considered. In our ISSV model, a single solution with
shock is able to give a global description for ISM shell and outer
region around an interstellar bubble.
Naturally, our dynamic model with self-gravity is more realistic and
can indeed describe expanding voids around which static and flowing
ISM solutions exist outside an expanding shock front (Figs.
\ref{fig:void_SIS} and \ref{fig:betaii}; Tables 4 and 5).

For gas dynamics, another problem for the absence of gravity is
revealed by asymptotic solution (\ref{largeasymp}),
where the coefficient of $x^{-4}$ term in $\alpha(x)$ differs by a
factor of $1/2$ between models with and without self-gravity, and
the expression for $v(x)$ at large radii differs from the $x^{-1}$
term.
These differences lead to different dynamic evolutions of voids (see
Section 2.1 for details).

\begin{figure} \centering
\includegraphics[width=0.5\textwidth]{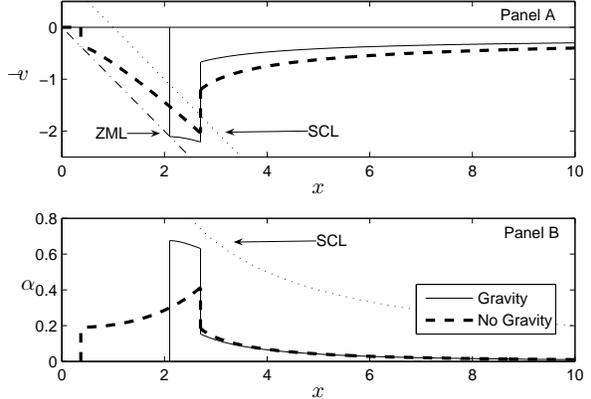}
\caption{
A comparison of ISSV solutions from different ODEs. Panel A presents
$-v(x)$ versus $x$ profiles and panel B presents $\alpha(x)$ versus
$x$ profiles. The light solid curves is a solution of ODEs
(\ref{ODE1}) and (\ref{ODE2}) with self-gravity. The heavy dashed
curves represent a solution of ODEs (\ref{ODE1_noG}) and
(\ref{ODE2_noG}) without self-gravity. An equi-temperature
shock with $\tau=1$ is introduced in both solutions at
$x_s=2.7025$. The dotted curves in both panels represent the SCL.
Other relevant parameters are contained in Table 6.
}\label{fig:Gravity}
\end{figure}

Under certain circumstances, the subtle difference between the shell
behaviours with and without gas self-gravity may result in quite
different shell profiles around void regions. We illustrate an
example for such differences
in Fig. \ref{fig:Gravity} with relevant parameters for the two
solutions therein
being summarized in Table 6. Given the same asymptotic condition of
$V=0.2$ and $A=1$ at large radii (Chevalier 1997a shows only $A=1$
case), the behaviours of such voids differ from each other
significantly with and without self-gravity, although both of them
fit asymptotic condition (\ref{largeasymp}) well. With gravity, the
given boundary condition and shock wave lead to a thin, very dense
shell with a sharp void edge while the same condition leads to a
quasi-smooth edge void
without gravity.
In Section 4.2, we will show that these two types of voids reveal
different processes to generate and maintain them. In certain
situations, void models without gravity might be misleading.

\begin{table}
\tabcolsep 0pt \caption{Parameters for two ISSV solutions shown in
Fig. \ref{fig:Gravity}}\vspace*{-12pt}
\begin{center}
\def\temptablewidth{0.45\textwidth}
{\rule{\temptablewidth}{1pt}}
\begin{tabular*}{\temptablewidth}{@{\extracolsep{\fill}}cccccc}
Gravity & $x_0$ &$\alpha_0$  &  $x_s$ & $V$ &  $A$ \\ \hline
With    & $2.1$ &  $0.675$   &$2.7025$&$0.2$& $1$\\
Without & $0.37$&  $0.191$   &$2.7025$&$0.2$& $1$\\
       \end{tabular*}
       {\rule{\temptablewidth}{1pt}}
       \end{center}
       \end{table}

\subsection[]{Formation of ISSV Edge}

In our model, the central void region is simply treated as a vacuum
with no materials inside.
We need to describe astrophysical mechanisms responsible for
creating such voids and for their local evolution. On the right side
of the void edge $x_0^{+}$, the gas density $\alpha(x)>0$ for
$x>x_0$, while $\alpha(x)=0$ for $x<x_0$. If no materials exist
within the void edge, there would be no mechanism to confine the gas
against the inward pressure force across the void edge. We offer two
plausible astrophysical scenarios to generate and maintain such
voids.

\subsubsection[]{Energetics and Pressure Balance}

If we allow a tenuous gas to exist within a `void' region to
counterbalance the pressure across the `void' edge for a certain
time $t$,
transformation (\ref{ST}) gives the edge density $\rho_0$ as
\begin{equation}\label{r1}
  \rho_0(t)=7.16\times10^{8}\alpha_0\left(\frac{10^3
  \hbox{ yr}}{t}\right)^2 m_p\ \ \hbox{cm}^{-3},
\end{equation}
where $\alpha_0\equiv\alpha(x_0)$ is the reduced mass density at
ISSV edge and $m_p$ is the proton mass.
For a gas temperature $T$,
the isothermal sound speed $a$ is
\begin{equation}\label{r2}
  a=2.87\times10^7\bigg(\frac{T}{10^7\hbox{ K}}\bigg)^{1/2}\ \hbox{cm
  s}^{-1},
\end{equation}
where the mean particle mass is that of hydrogen atom.
Then the gas pressure $p_0$ just on the outer side of the ISSV edge
is
\begin{equation}\label{r3}
 p_0(t)=\rho_0a^2=0.99\alpha_0
 \left(\frac{10^3\hbox{ yr}}{t}\right)^2
 \frac{T}{10^7\hbox{ K}}\hbox{ dyne cm}^{-2},
\end{equation}
where $\rho_0$ is the mass density at the ISSV edge.
Here, we take the proton mass as the mean particle mass. Equation
(\ref{r3}) gives a pressure scaling $p_0\propto t^{-2}T$ governed by
self-similar hydrodynamics.

Within a `void', we may consider that a stellar wind steadily blows
gas outwards with a constant speed. Various astrophysical systems
can release energies at different epochs.
For an early evolution, massive stars steadily blow strong winds
into the surrounding ISM (e.g. Mathews 1966; Dyson 1975; Falle 1975;
Castor et al. 1975; Weaver et al. 1977).
In the late stage of evolution, compact stars can also blow fast
winds to drive the surrounding gas (e.g. Chevalier 1997a; Chevalier
1997b). As a model simplification, we assume that a tenuous gas
moving outwards at constant speed $v_w$ with temperature $T_w$ (we
refer to this as a central wind) may carve out a central `void' and
can provide a pressure against the pressure gradient across the
`void' boundary; suppose that this central wind begins to blow at
time $t=0$. Then after a time $t$, the radius $r$ of the central
wind front is at $r_w=v_wt$. By the mass conservation in a
spherically symmetric flow, the mass density of the central wind
front at time $t$ is then
\begin{equation}\label{r5}
 \rho_{w, front}=\frac{\dot{M}}{4\pi v_w^3t^2}\ ,
\end{equation}
where $\dot{M}$ is the mass loss rate of the central wind.
For a contact discontinuity the ISSV edge between the inner stellar
wind front and outer slower wind,
the plasma pressure $p_{w, front}$ of this central fast wind is
\begin{equation}\label{r6}
 p_{w, front}=\frac{k_BT_w}{m}\frac{\dot{M}}{4\pi v_w^3t^2}\ ,
\end{equation}
which can be estimated by
\begin{equation}\label{r7}
\frac{4.16\times10^{-6}\dot{M}}{10^{-6}M_{\odot}\hbox{ yr}^{-1}}
\left(\frac{10\hbox{km s}^{-1}}{v_w}\right)^3\left(\frac{10^3
\hbox{yr}}{t}\right)^2\frac{T_w}{10^7\hbox{K}}\hbox{dyne\ cm}^{-2},
\end{equation}
where $M_{\odot}$ is the solar mass and the mean particle mass is
that of hydrogen atom. We adopt the parameters based on estimates
and numerical calculations (see Section 4.3).
By expressions (\ref{r6}) and (\ref{r7}), the plasma pressure at the
central wind front also scales as $p_{w, front}\propto t^{-2}T_w$.
By a contact discontinuity between the central wind front and the
ISSV edge with a pressure balance $p_{w, front}=p_0$, our
self-similar `void' plus the steady central wind can sustain an ISSV
evolution as long as the central stellar wind can be maintained.
Across such a contact discontinuity, the densities and temperatures
can be different on both sides.
Equations (\ref{r7}) and (\ref{r3}) also show that, while pressures
balance across a contact discontinuity, the reduced density at ISSV
edge $\alpha_0$ is determined by the mass loss rate $\dot M$ and
central wind radial velocity $v_w$.
Back into this steady central stellar wind of a tenuous plasma at a
smaller radius, it is possible to develope a spherical reverse
shock. This would imply an even faster inner wind inside the reverse
shock (i.e. closer to the central star); between the reverse shock
and the contact discontinuity is a reverse shock heated downstream
part of the central stellar wind. Physically, the downstream portion
of the central stellar wind enclosed within the contact
discontinuity is expected to be denser, hotter and slower as
compared with the upstream portion of the central stellar wind
enclosed within the reverse shock.

The above scenario may be also adapted to supernova explosions.
At the onset of supernova explosions, the flux of energetic
neutrinos generated by the core collapse of a massive progenitor
star could be the main mechanism to drive the explosion and
outflows.
This neutrino pressure, while different from the central wind plasma
pressure discussed above, may be able to supply sufficient energy to
drive rebound shocks in a dense medium that trigger supernova
explosions (e.g. Chevalier 1997b; Janka et al. 2007; Lou \& Wang
2006, 2007; Lou \& Cao 2007; Arcones et al. 2008; Hu \& Lou 2009).
Under certain conditions, such neutrino pressure may even
counterbalance the strong inward pressure force of an extremely
dense gas and generate central `voids'.
We will apply this scenario and give
examples in Section 4.3.

\subsubsection[]{Diffusion Effects}

For astrophysical void systems with timescales of sufficient energy
supply or pressure support being shorter than their ages, there
would be not enough outward pressure at void edges to balance the
inward pressure of the gas shell surrounding voids after a certain
time. In such situations, the gas shell surrounding a central void
will inevitably diffuse into the void region across the void edge or
boundary. This diffusion effect will affect behaviours of void
evolution especially in the environs of void edge and gradually
smear the `void boundary'.
However, because the gas shell
has already gained a steady outward radial velocity before the
central energy supply such as a stellar wind pressure support fails,
the inertia of a dense gas shell will continue to maintain a shell
expansion for some time during which the gas that diffuses into the
void region accounts for a small fraction of the entire shell and
outer envelope. As a result, it is expected that our ISSV solutions
remain almost valid globally to describe void shell behaviours even
after a fairly long time of insufficient central support.

We now estimate the gas diffusion effect quantitatively. We assume
that a void boundary expands at $ax_0$
and the void is surrounded by a gas envelope whose density profile
follows a $\rho(r)\propto r^{-2}$ fall-off (asymptotic solution
\ref{largeasymp}). The gas shell expands at radial velocity $ax$ for
$x>x_0$. The central energy supply mechanism has already maintained
the void to a radius $r_0$ and then fails to resist the inward
pressure across the void edge. We now estimate how many gas
particles diffuse into void region in a time interval of $\Delta
t=r_0/(ax_0)$, during which the void is supposed to expand to a
radius $2r_0$. We erect a local Cartesian coordinate system in an
arbitrary volume element in gas shell with the $x-$axis pointing
radially outwards. The Maxwellian velocity distribution of thermal
particles gives the probability density of velocity
$\overrightarrow{v}=(v_x, v_y, v_z)$ as
\begin{equation}\label{r8}
p_v(\overrightarrow{v})\propto
\exp{\left[-\frac{(v_x-ax_0)^2+v_y^2+v_z^2}{2a^2}\right]}\ .
\end{equation}
Define $l$ as the mean free path of particles in the gas shell near
void edge. If a gas particle at radius $r$ can diffuse into radius
$\tilde{r}$ without collisions, its velocity is limited by
$(r+v_x\Delta t)^2+(v_y\Delta t)^2+(v_z\Delta t)^2<\tilde{r}^2$ and
its position is limited by $r-\tilde{r}<l$. We simplify the velocity
limitation by slightly increasing the interval as $(r+v_x\Delta
t)^2<\tilde{r}^2$ and $v_y^2+v_z^2<(\tilde{r}/\Delta t)^2$. We first
set $\tilde{r}=r_0$ and integrate the ratio of particles that
diffuse into radius $r_0$ during $\Delta t$ to total gas shell
particles within radius $r_0+l$ as
\begin{multline}\label{r9}
\!\!\!\!
\frac{\int_{r_0}^{r_0+l}4\pi
r^2\rho(r)dr\int_{\|r+v_x\Delta t\|<r_0\&v_y^2+v_z^2<(r_0/\Delta
t)^2} p_vd^3v}
{\int_{r_0}^{r_0+l}4\pi r^2\rho(r)dr\int p_vd^3v}\\
=\frac{1-\exp{\left[-r_0^2/(2a^2\Delta
t^2)\right]}}{(2\pi)^{1/2}l}\\
\qquad\qquad\times\int_{r_0}^{r_0+l}dr\int_{-\frac{r+r_0}{a\Delta
t}-x_0}^{-\frac{r-r_0}{a\Delta t}-x_0}\exp{(-\tilde{v}^2/2)}d\tilde{v}\\
=\frac{1-\exp{(-x_0^2/2)}}{(2\pi)^{1/2}}\frac{r_0}{l}\\
\qquad\ \times\int_1^{1+l/r_0}
d\tilde{x}\int_{-\tilde{x}-2x_0}^{-\tilde{x}}
\exp{\left(-\tilde{v}^2/2\right)}d\tilde{v}\ ,
\end{multline}
where both $\tilde{x}$ and $\tilde{v}$ are integral elements. We
simply set $x_0=1$ and present computational results in Table 7. It
is clear that even the inner energy supply fails to sustain inward
pressure for a fairly long time, there are only very few particles
that diffuse into the original void region, namely, a void remains
quite empty.

In the context of PNe,
the particle mean free path $l$ may be estimated for different
species under various situations. For an example of PN to be
discussed in Section 4.3.1, $l=1/(n\sigma)=3\times10^{20}$ cm,
where $n\approx 5000$ cm$^{-3}$ is the proton (electron) number
density in the H II region and $\sigma=6.65\times 10^{-25}$ cm$^2$
is the electron cross section
in Thomson scattering.
One can also estimate cross section of coulomb interaction between
two protons as $\sim 10^{-17}\hbox{ cm}^2$ and thus the mean free
path for proton collisions is $\sim 10^13$cm. A PN void radius is
$\sim 5\times10^{17}$ cm. If no inner pressure is acting further
as a void expands to radius $\sim10^{18}$ cm, gas particles that
diffuse into $r\lsim 5\times10^{17}$ cm only take up
$\sim6\times10^{-5}$ of those in the gas shell. However, at the
onset of a supernova explosion, the particle mean free path in the
stellar interior is very small. If the density at a void edge is
$\sim 1.2\times10^8$ g cm$^{-3}$ (see Section 4.3.2),
and the scattering cross section is estimated as the iron atom cross
section $4.7\times 10^{-18}$ cm$^2$, the particle mean free path is
only $l\sim 10^{-13}$ cm. Under this condition, particle that
diffuse into a void region can account for $6.2\%$ of those in the
total gas shell.

In short, while diffusion effect inevitably occurs when the inner
pressure can no longer resist the inward gas pressure across a void
edge, it usually only affects the gas behaviour near the void edge
but does not alter the global dynamic evolution of gas shells and
outer envelopes over a long time. However, we note that for a very
long-term evolution, there will be more and more particles
re-entering the void region when no sufficient pressure is supplied,
and eventually the diffusion effect will result in significant
changes of global dynamical behaviours of voids and shells.
These processes may happen after supernova explosions. When
neutrinos are no longer produced and rebound shocks are not strong
enough to drive outflows, a central void generated in the
explosion will gradually shrink in the long-term evolution of
supernova remnants (SNRs).

\begin{table}
\tabcolsep 0pt \caption{Ratio of molecules that diffuse into
radius $r_0$ during $\Delta t$ to total gas shell molecules within
radius $r_0+l$ under different ratio $l/r_0$. We take $x_0=1$.}
\vspace*{-12pt}
\begin{center}
\def\temptablewidth{0.45\textwidth}
{\rule{\temptablewidth}{1pt}}
\begin{tabular*}{\temptablewidth}{@{\extracolsep{\fill}}cccccc}
$l/r_0$ & $+\infty$ & $10$    &  $1$   & $0.1$ &  $<<1$\\
\hline
ratio   & $3.6\%r_0/l$ & $0.34\%$& $2.9\%$& $5.8\%$& $6.2\%$\\
       \end{tabular*}
       {\rule{\temptablewidth}{1pt}}
       \end{center}
       \end{table}

\subsubsection[]{Applications of ISSV Model Solutions}

In formulating the basic model, we ignore the gravity of the
central void region. By exploring the physics around the void
boundary, a tenuous gas is unavoidable inside a `void' region for
either an energy supply mechanism leading to an effective pressure
or that diffused across the void boundary.
As long as the gravity associated with such a tenuous gas inside
the `void' is sufficiently weak, our ISSV model should remain
valid for describing the large-scale dynamic evolution of void
shells, shocks, and outer envelopes.

In a planetary nebula or a supernova remnant, there is usually a
solar mass compact star at the centre (e.g. Chevalier 1997a).
For outgoing shells at a slow velocity of sound speed $\sim 10$ km
s$^{-1}$,
the Parker-Bondi radius of a central star of $10M_{\odot}$ is $\sim
10^{15}$ cm (i.e. $\sim10^{-3}$ light year or $\sim 3\times 10^{-4}$
pc). The typical radius of a planetary nebula is about $10^{18}$ cm
(see Section 4.3.1). Even the youngest known supernova remnant
G1.9+0.3 in our Galaxy, estimated to be born $\sim 140$ yr ago,
has a radius of $\sim 2$ pc (e.g. Reynolds 2008; Green et al.
2008). Thus the central star only affects its nearby materials and
has little impact on gaseous remnant shells.

For a stellar wind bubble (e.g. Rosette Nebula), there usually are
several, dozens or even thousands early-type stars blowing strong
stellar winds in all directions. For example, the central `void'
region of Rosette Nebula contains the stellar cluster NGC 2244 of
$\sim 2000$ stars (e.g. Wang et al. 2008). Conventional estimates
show that the thick nebular shell has a much large mass of around
$10$,$000-16$,$000$ solar masses (e.g. Menon 1962; Krymkin 1978).
For a sound speed of $\sim 10$ km s$^{-1}$, the Parker-Bondi radius
of a central object of $2000M_{\odot}$ is then $\sim 0.08$ pc, which
is again very small compared to the $\sim 6$ pc central void in
Rosette Nebula (e.g. Tsivilev et al. 2002). For a typical
interstellar bubble consdiered in Weaver et al. (1977), the total
mass inside the bubble, say, inside the dense shell, is no more than
$50$ solar masses, which is significantly lower than that of the
$2000$ solar mass dense shell.

We thus see that the dynamical evolution of flow systems on scales
of planetary nebulae, supernova remnants and interstellar bubbles
are only affected very slightly by central stellar mass objects
(e.g. early-type stars, white dwarfs, neutron stars etc).
Based on this consideration, we regard the grossly spherical
region inside the outer dense shells in those astrophysical systems
as a void in our model formulation,
ignore the void gravity, emphasize the shell self-gravity and invoke
the ISSV solutions to describe their dynamic evolution.

\subsection[]{Astrophysical Applications}

Our ISSV model is adaptable to astrophysical flow systems such as
planetary nebulae, supernova explosions, supernova remnants, bubbles
and hot bubbles on different scales.

\subsubsection[]{Planetary Nebulae}

In the late phase of stellar evolution, a star with a main-sequence
mass $\lsim 8M_{\odot}$ makes a transition from an extended, cool
state where it blows a slow dense wind to a compact, hot
state\footnote{A hottest white dwarf (KPD 0005 5106) detected
recently has a temperature of $\sim 2\times 10^5$ K (Werner et al.
2008).} where it blows a fast wind. The interaction between the
central fast wind with the outer slow wind results in a dense shell
crowding the central region which appears as a planetary nebula
(e.g. Kwok, Purton \& Fitzgerald 1978). The hot compact white dwarf
star at the centre is a source of photoionizing radiation to ionize
the dense shell (e.g. Chevalier 1997a). When the fast wind catches
up the slow dense wind, a forward shock and a reverse shock will
emerge on outer and inner sides of a contact discontinuity,
respectively.
As shown in Section 4.2.3, the gravity of central white dwarf and
its fast steady wind is negligible for the outer dense wind.
Practically, the region inside the contact discontinuity may be
regarded approximately as a void.
Meanwhile, the
photoionizing flux is assumed to be capable of ionizing and
heating the slow wind shell to a constant temperature (Chevalier
1997a), and the outer envelope, the cool AGB slow wind that is
little affected by the central wind and radiation, can be also
regarded approximately as isothermal. The constant temperatures of
dense photoionized shell and outer envelope are usually different
from each other which can be well characterized by the isothermal
sound speed ratio $\tau$ (see Section 2.2). Thus the dynamic
evolution of dense shell and outer AGB wind envelope separated by
a forward shock is described by a type
$\mathcal{Z}$ ISSV solution.

Within the contact discontinuity spherical surface, there is a
steady downstream wind blowing outside the reverse shock front
(e.g. Chevalier \& Imamura 1983). Consistent with Section 4.2.1,
we define $r_w$ as the radius where the downstream wind front
reaches and $r_r$ as the radius of reverse shock. Within the
radial range of $r_r<r<r_w$, i.e. in the downstream region of the
reverse shock, we have a wind mass density
\begin{equation}\label{fastwind_density}
 \rho_w(r,\ t)=\frac{\dot{M}}{4\pi v_wr^2}\ .
\end{equation}
%
We define $a_{w,d(u)}$ and $T_{w,d(u)}$ as the sound speed and gas
temperature on the downstream (upstream) side of the reverse shock
and ratio $\tau_w\equiv a_{w,d}/a_{w,u}$
$=(T_{w,d}/T_{w,u})^{1/2}$ to characterize the reverse shock. For
a reverse shock in the laboratory framework of reference given by
shock conditions (\ref{shockcm}) and (\ref{shockcmom}), we have in
dimensional forms
%
\begin{multline}\label{general_reverse_shock}
\!\!\!\!\!\!\! u_{u,rs}-u_{rs}=\frac{1}{2}\left(v_w-u_{rs}
+\frac{a_{w,d}^2}{v_w-u_{rs}}\right)\\
\!\!\!
+\frac{1}{2}\left\lbrace\left[\frac{(v_w-u_{rs})^2
-a_{w,d}^2}{v_w-u_{rs}}\right]^2
+4a_{w,d}^2\frac{\tau_w^2-1}{\tau_w^2}\right\rbrace^{1/2}\!\! ,\\
\!\!\!\!\!\!\!
v_w-u_{rs}=\frac{1}{2}\left(u_{u,rs}-u_{rs}+\frac{a_{w,u}^2}{u_{u,rs}
-u_{rs}}\right)\\
\!\!\!
-\frac{1}{2}\left\lbrace\left[\frac{(u_{u,rs}-u_{rs})^2
-a_{w,u}^2}{u_{u,rs}-u_{rs}}\right]^2
+4a_{w,u}^2(1-\tau_w^2)\right\rbrace^{1/2}\!\! ,\\
\!\!\!\!\!\!\!
\rho_{u,rs}=\frac{v_w-u_{rs}}{u_{u,rs}-u_{rs}}\rho_{d,rs}\ ,\\
\end{multline}
where $u_{rs}$ is the outgoing speed of the reverse shock,
$u_{u,rs}$ is the upstream wind velocity,  $\rho_{u(d), rs}$ is
the upstream (downstream) mass density, respectively. The first
and second expressions in equation (\ref{general_reverse_shock})
are equivalent: the first expresses the upstream flow velocity in
terms of the downstream parameters, while the second expresses the
downstream flow velocity in terms of the upstream parameters.
In solving the quadratic equation, we have chosen the physical
solution, while the unphysical one is abandoned. Normally, in the
downstream region $r_r<r<r_w$, the plasma is shock heated by the
central faster wind with $\tau_w>1$. In the regime of an
isothermal shock for effective plasma heating, we take $\tau_w=1$
(i.e. $a_{w,d}=a_{w,u}=a_w$) for a stationary reverse shock in the
laboratory framework of reference, shock conditions
(\ref{general_reverse_shock}) and (\ref{fastwind_density}) reduce
to
\begin{equation}\label{static_reverse_shock}
\rho_{u, rs}=\frac{\dot{M}}{4\pi u_{u, rs}r_r^2}\ ,\qquad u_{u,
rs}=\frac{a_w^2}{v_w}\ ,\qquad u_{rs}=0\ .
\end{equation}
%
In this situation, the reverse shock remains stationary in space
and this may shed light on the situation that an inner fast wind
encounters an outer dense shell of slow speed.
While a
reverse shock always moves inwards relative to both upstream and
downstream winds, either outgoing or incoming reverse shocks are
physically allowed in the inner wind zone in the laboratory
framework of reference. In the former situation, the reverse shock
surface, contact discontinuity surface and forward shock surface
all expand outwards steadily, with increasing travelling speeds,
respectively. In the latter situation, the downstream wind zone,
namely, the shocked hot fast wind zone expands both outwards and
inwards, and eventually fills almost the entire spherical volume
within the dense shell. The reverse shock here plays the key role
to heat the gas confined within a planetary nebula to a high
temperature, which is thus referred to as a hot bubble.
In all situations, between the reverse shock and the contact
discontinuity, the downstream wind has a constant speed, supplying a
wind plasma pressure to counterbalance the inward pressure force
across the contact discontinuity.

Here, type $\mathcal{Z}$ ISSV solutions are utilized to describe
the self-similar dynamic evolution of gas shell outside the
outgoing contact discontinuity. An outgoing forward shock
propagates in the gas shell after the central fast wind hits the
outer dense shell. According to properties of type $\mathcal{Z}$
ISSV solutions, there may be outflows (i.e. winds and breezes),
static ISM or inflows (i.e. accretion flows and contractions) in
the region outside the forward shock. The spatial region between
the contact discontinuity and the forward shock is the downstream
side of the forward shock.

%

\begin{figure} \centering
\includegraphics[width=0.5\textwidth]{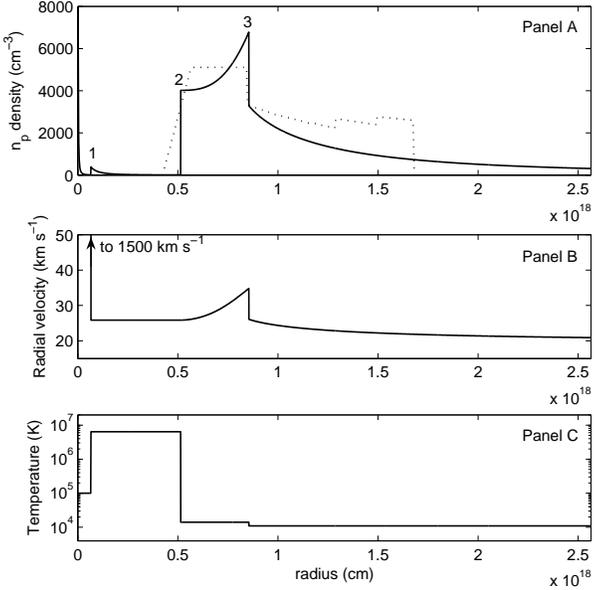}
\caption{A type $\mathcal{Z}$ ISSV shock model with an inner fast
wind to fit the planetary nebula NGC 7662. Panels A, B and C show
our model results for proton number density $n_{\rm p}$, radial flow
velocity and temperature of NGC 7662 in solid curve, respectively.
Numerals $1$, $2$ and $3$ in Panel A mark reverse shock, contact
discontinuity and forward shock surface, respectively. The dashed
curve in Panel A is the estimate of proton number density by
Guerrero et al. (2004). The inner fast wind blows at $1500$ km
s$^{-1}$ inside the reverse shock, which is not shown in Panel
B.}\label{fig:NGC7662}
\end{figure}

We now provide our quantitative model PN estimates for a
comparison. Guerrero et al. (2004) probed the structure and
kinematics of the triple-shell planetary nebula NGC 7662 based on
long-slit echelle spectroscopic observations and Hubble Space
Telescope archival narrowband images. They inferred that the
nebula with a spatial size of $\sim 4\times10^{18}$ cm consists of
a central cavity surrounded by two concentric shells, i.e. the
inner and outer shells, and gave a number density $n_{\rm p}$
distribution (the dotted curve in panel A of Figure
\ref{fig:NGC7662}). The temperatures of the inner and outer shells
were estimated as $\sim 1.4\times 10^4$ K and $\sim 1.1\times
10^4$ K, respectively. No information about the inner fast wind is
given in Guerrero et al. (2004). In our model consideration, the
planetary nebula NGC 7662 may be described by our type
$\mathcal{Z}$ ISSV model with a shocked inner fast wind. In our
scenario for a PN, the central cavity in the model of Guerrero et
al. (2004) should actually involve an inner fast wind region with
a reverse shock. The inner and outer shells correspond to the
downstream and upstream dense wind regions across a forward shock,
respectively. Thus the inner boundary of the inner shell is the
contact discontinuity in our model scenario. Physically, we
suppose that the central star stops to blow a dense slow wind of
$\sim 10$ km s$^{-1}$ about $\sim 1000$ years ago and the inner
fast wind of $10^5$ K began to blow outwards from a white dwarf at
$u_{u,rs}=1500$ km s$^{-1}$ about $\sim 600$ years ago. When the
inner fast wind hits the dense slow wind $\sim 4$ years after its
initiation, a reverse shock and a forward shock are generated on
the two sides of the contact discontinuity. The reverse shock
moves inwards at a speed $\sim 10$ km s$^{-1}$. The best density
fit to the estimate of Guerrero et al. (2004) is shown in Figure
\ref{fig:NGC7662}. In our model, the inner fast wind has a mass
loss rate from the compact star as $\sim 2\times
10^{-8}$M$_{\odot}$ yr$^{-1}$ (consistent with earlier estimates
of Mellema 1994 and Perinotto et al. 2004) and the reverse shock
is able to heat the downstream wind to a temperature of $\sim
6.4\times 10^6$ K. The downstream wind of the reverse shock has an
outward speed of $\sim 25.9$ km s$^{-1}$, corresponding to a
kinematic age (i.e. the time that a shocked fast wind at that
velocity blows from the central point to its current position) of
$\sim 630$ years. In Guerrero et al. (2004), the kinematic age is
estimated to be $\sim 700$ years. In Guerrero et al. (2004), the
inner shell density is $\sim 5\times 10^3$ $m_{\rm p}$ cm$^{-3}$
and our model shows a density variation from $\sim4\times10^3$
$m_{\rm p}$ cm$^{-3}$ to $\sim7\times10^3$ $m_{\rm p}$ cm$^{-3}$
with a comparable mean. The forward shock travels outwards at a
speed $\sim 43.0$ km s$^{-1}$, consistent with an average outward
velocity of the inner shell at $\sim 44$ km s$^{-1}$. The total
mass of the inner shell is $\sim 8.5\times10^{-3}$M$_{\odot}$, and
the mass of the outer shell within a radius of $2.5\times10^{18}$
cm is $\sim 0.036$M$_{\odot}$, which are all consistent with
estimates of Guerrero et al. (2004). However, Guerrero et al.
inferred that the outer shell has an outward velocity of around
$50$ km s$^{-1}$ and a proton number density of $\sim 3000$
cm$^{-3}$. Our model estimates indicate that the outer shell has a
proton number density from $3200$ cm$^{-3}$ at the immediate
upstream side of the forward shock to $\sim 400$ cm$^{-3}$ at
$2.5\times 10^{18}$ cm, and the outward velocity varies from $\sim
26$ km s$^{-1}$ at the upstream point of the forward shock to
$\sim 20$ km s$^{-1}$ at $2.5\times 10^{18}$ cm. And thus the
dense slow wind mass loss rate is
$0.68\times10^{-5}$M$_{\odot}$yr$^{-1}$, which is consistent with
earlier numerical simulations (e.g. Mellema 1994; Perinotto et al.
2004), but is lower by one order of magnitude than $\sim
10^{-4}$M$_{\odot}$yr$^{-1}$ estimated by Guerrero et al. (2004).
In summary,
our ISSV model appears consistent with observations of the NGC
7662, and a combination of hydrodynamic model with optical and
X-ray observations would be valuable to understand the structure
and dynamic evolution of planetary nebulae.

\subsubsection[]{Supernova Explosions and Supernova Remnants}

At the onset of a type II supernova (or core-collapse supernova)
for a massive progenitor, extremely energetic neutrinos are
emitted by the neutronization process to form a `neutrino sphere'
that is deeply trapped by the nuclear-density core and may trigger
a powerful rebound shock breaking through the heavy stellar
envelope. At that moment, the central iron core density of a $\sim
15M_{\odot}$ progenitor star can reach as high as $\sim
7.3\times10^9$ g cm$^{-3}$ and the core temperature could be
higher than $\sim 7.1\times10^9$ K. The density of the silicon
layer is $\sim 4.8\times10^7$ g cm$^{-3}$ with a temperature of
$\sim 3.3\times 10^9$ K. The tremendous pressure produced by
relativistic neutrinos may drive materials of such high density to
explosion (e.g. Woosley \& Janka 2005).
During the first $\sim 10$ s of the core collapse, a power of about
$\sim 10^{53}$ erg s$^{-1}$ is released as high-energy neutrinos
within a radius of $\sim 10^5$ km (e.g. Woosley \& Janka 2005). The
neutrino-electron cross section was estimated to be $\sim
10^{-42}$(E/GeV)cm$^2$ with $E$ being the neutrino energy (e.g.
Marciano \& Parsa 2003). During the gravitational core collapse of a
SN explosion, a typical value of E would be E$\sim $20MeV (e.g.
Hirata et al. 1987; Arcones et al. 2008). Therefore, the
neutrino-electron cross section is estimated to be $\sim 2\times
10^{-44}\hbox{ cm}^2$. If we adopt the neutrino luminosity
$L=10^{53}\hbox{ erg s}^{-1}$, the ratio of neutrino pressure to one
electron and the iron-core gravity on one silicon nucleus is $\sim
10^{-6}$. By these estimates, neutrino pressure is unable to split
the silicon layer from the iron core. A pure vacuum void is unlikely
to appear during the gravitational core collapse and the rebound
process to initiate a SN.

During the subsequent dynamic evolution, diffusion effects would
gradually smooth out any sharp edges
and the inner rarefied region will be dispersed with diffused
gaseous materials. Behaviour of this gas may be affected by the
central neutron star. For example, a relativistic pulsar wind is
able to power a synchrotron nebula referred to as the pulsar wind
nebula, which is found within shells of supernova remnants (e.g.
Gaensler et al. 1999). Pulsar winds relate to the magnetic field of
pulsars and are not spherically symmetric. However, if the angle
between magnetic and spin axes of a pulsar is sufficiently large and
the pulsar is rapidly spinning, the averaged pressure caused by a
pulsar wind may appear grossly spherical. This pulsar wind pressure
may also counterbalance the inward pressure of outer gas and slow
down the diffusion process. We offer a scenario that a supernova
remnant makes a transition from a sharp-edged `void' to a
quasi-smooth one due to combined effects of diffusion and pulsar
wind.


\subsubsection[]{Interstellar Bubbles}

Our ISSV model solutions may also describe large-scale nebula
evolution around early-type stars
or Wolf-Rayet stars. Here
our overall scenario parallels to the one outlined in Section 4.3.1
for PNe but is different at the ambient medium surrounding the flow
systems. For an interstellar bubble, a central stellar wind collides
with the ISM (i.e. no longer a dense wind) and gives rise to a
reverse shock that heats the downstream stellar wind zone, and a
forward shock that propagates outwards in the ISM. Meanwhile, by
inferences of radio and optical observations for such nebulae (e.g.
Carina Nebula and Rosette Nebula), the central hot stars are capable
of ionizing the entire swept-up shell and thus produce huge H II
regions surrounding them (e.g. Menon 1962; Gouguenheim \& Bottinelli
1964; Dickel 1974). The temperature of a H II region is usually
regarded as weakly dependent on plasma density, thus an isothermal H
II region should be a fairly good approximation (e.g. Wilson, Rohlfs
\& H\"uttemeister 2008). As the ISM outside the forward shock is
almost unaffected by the central wind zone, we may approximate the
static ISM as isothermal and the dynamic evolution of gas
surrounding interstellar bubbles can be well characterized by
various ISSV solutions. Several prior models that describe ISM
bubble shells in adiabatic expansions without gravity, might
encounter problems. For example, the self-similar solution of Weaver
et al. (1977) predicts a dense shell with a thickness of only $\sim
0.14$ times the radial distance from the central star to the shell
boundary, which is indeed a very thin shell. However, observations
have actually revealed many ISM bubbles with much thicker shells
(e.g. Dorland, Montmerle \& Doom 1986). The Rosette Nebula has a H
II shell with a thickness of $\sim 20$ pc while the radius of
central void is only $\sim 6$ pc (e.g. Tsivilev et al. 2002). The
shell thickness thus accounts for $\sim 70\%$ of the radius of shell
outer boundary, which is much larger than the computational result
of Weaver et al. (1977). In our ISSV solutions, there are more
diverse dynamic behaviours of gas shells and outer envelopes. For
example in Figure 9, four ISSV solutions with static outer envelope
indicate that the ratio of shell thickness to the radius of the
forward shock covers a wide range. For these four solutions, this
ratio is $0.65$, $0.98$, $0.91$ and $0.57$ respectively.
A type $\mathcal{Z}_I$ void shock solution with
$x_0=0.58$, $\alpha_0=0.014$ and an isothermal shock at $x_s=2.2$
may characterize gross features of Rosette Nebula reasonably well.
This ISSV solution indicates that when the constant shell
temperature is $7000$ K (somewhat hotter than $6400$ K as inferred
by Tsivilev et al. 2002) and the entire nebula system has evolved
for $\sim 10^6$ years, the central void has a radius of $\sim 6.0$
pc; the forward shock outlines the outer shell radius as $\sim
22.8$ pc; the electron number density in the HII shell varies from
$\sim 8.5$ cm$^{-3}$ to $\sim 12.9$ cm$^{-3}$; the contact
discontinuity surface expands at a speed of $\sim 6.1$ km s$^{-1}$
and the forward shock propagates into the ISM at a speed of $\sim
16.4$ km s$^{-1}$; the surrounding ISM remains static at large
radii. In the above model calculation, an abundance He$/$H ratio
of $0.1/0.9$ is adopted. Various observations lend supports to our
ISSV model results. For example, Tsivilev et al. (2002) estimated
a bit higher shell electron number density as $15.3$ cm$^{-3}$ and
an average shell expanding velocity of about $8.5$ km s$^{-1}$.
Dorland et al. (1986) gave an average shell electron number
density as $11.3$ cm$^{-3}$. Our calculation also gives a shell
mass of $\sim 1.55\times 10^4M_{\odot}$, which falls within
$10$,$000M_{\odot}$ and $16$,$000M_{\odot}$ as estimated by Menon
(1962) and Krymkin (1978), respectively.



To study these inner voids embedded in various gas nebulae,
diagnostics of X$-$ray emissions offers a feasible means to probe
hot winds. The thermal bremsstrahlung and line cooling mechanisms
can give rise to detectable X$-$ray radiation from optically thin
hot gas (e.g. Sarazin 1986) and the high temperature interaction
fronts of stellar winds with the ISM and inner fast wind with the
void edge can produce X$-$ray photons (e.g. Chevalier 1997b). We
will provide the observational properties of different types of
voids and present diagnostics to distinguish the ISSV types and
thus reveal possible mechanisms to generate and maintain such
voids by observational inferences in a companion paper (Lou \&
Zhai 2009 in preparation).

\section[]{Conclusions}

We have explored self-similar hydrodynamics of an isothermal
self-gravitating gas with spherical symmetry and shown various
void solutions without or with shocks.

We first obtain type $\mathcal{X}$ ISSV solutions without shocks
outside central voids in Section 3.3. Based on different
behaviours of eigen-derivatives across the SCL, type $\mathcal{X}$
void solutions are further divided into two subtypes: types
$\mathcal{X}_{\rm I}$ and $\mathcal{X}_{\rm II}$ ISSV solutions.
All type $\mathcal{X}$ ISSV solutions are characterized by central
voids surrounded by very dense shells. Both types
$\mathcal{X}_{\rm I}$ and $\mathcal{X}_{\rm II}$ ISSV solutions
allow envelope outflows but only type $\mathcal{X}_{\rm I}$ ISSV
solutions can have outer envelopes in contraction or accretion
flows.

We then consider self-similar outgoing shocks in gas envelopes
surrounding central voids in Section 3.4. Type $\mathcal{Z}_{\rm
I}$ void solutions are referred to as the equi-temperature
shock solutions with a constant gas temperature across a shock
front (this is an idealization; see Spitzer 1978). We also
investigate various cases of shocks in type $\mathcal{Z}_{\rm II}$
void solutions (always a higher downstream temperature for the
increase of specific entropy). In Section 3.4, we have developed
the `phase net' matching procedure to search for type
$\mathcal{Z}$ ISSV solutions with static, expanding, contracting
and accreting outer envelopes. ISSV solutions with quasi-smooth
edges exist only when gas flows outwards outside the shock; all
other types of voids are surrounded by fairly dense shells or
envelopes.


We have systematically examined voids with sharp or quasi-smooth
edges for various ISSV solutions.
There must be some energetic processes such as supernova
explosions or powerful stellar winds (including magnetized
relativistic pulsar winds) that account for the appearance of
sharp-edge voids. The denser the nebular shell is, the more
difficult the void formation is. In other words, voids of
quasi-smooth edges might be easier to form in the sense of a less
stringent requirement for the initial energy input. In Section
4.1, we point out that the gas self-gravity can influence the void
evolution significantly, especially for the property of regions
near void edge as shown in Fig. \ref{fig:Gravity}. With the same
boundary condition of outer medium, ISSV models with and without
self-gravity can lead to different types of voids, e.g.
quasi-smooth edge or sharp-edge voids. We suggest that these two
types of voids may have different mechanisms to generate and
sustain. Thus the inclusion of gas self-gravity is both physically
realistic and essential. Besides, we show that all voids with
quasi-smooth edges are type $\mathcal{Z}$ voids, that is, with
shocks surrounding voids. This indicates that shock and expanding
outer envelope may well imply a likely presence of a central void.
In fact, observations on hot gas flows in clusters of galaxies
might also be relevant in this regard. For example, McNamara et
al. (2006) reported giant cavities and shock fronts in a distant
($z=0.22$) cluster of galaxies caused by an interaction between a
synchrotron radio source and the hot gas around. Such giant X-ray
cavities were reported to be left behind large-scale shocks in the
galaxy cluster MS0735.6+7421.


Another point to note is that ISSV solutions we have constructed are
physically plausible with special care taken for the expanding void
boundary $x_0$.
Void boundary $x_0$ involves density and velocity jumps
not in the sense of a shock; local diffusion processes should happen
to smooth out such jumps in a non-self-similar manner.
Nonlinear ODEs (\ref{ODE1}) and (\ref{ODE2}) are valid
in intervals $(0,\ x_0^{-})$ and $(x_0^{+},\ +\infty)$.
We have indicated this property in Section 3.2 when introducing
the concept of ISSV and discuss this issue in Section 4.2. There,
several plausible mechanisms are noted such as
powerful stellar winds and energetic neutrino driven supernova
explosion.
More specifically, we apply the ISSV solutions to grossly spherical
planetary nebulae, supernova explosions or supernova remnants and
interstellar bubbles.

Our model for planetary nebulae involve three characteristic
interfaces: reverse shock, contact discontinuity surface, and
forward shock. Steady inner stellar winds of different speeds blow
on both sides of the inner reverse shock and a contact discontinuity
surface confines the slower downstream wind zone outside the inner
reverse shock. This reverse shock may be stationary or moving
(either inwards or outwards) in the laboratory framework of
reference. The contact discontinuity surface between the steady
downstream wind zone (on the downstream side of the inner reverse
shock) and the outer expanding gas shell moves outwards at a
constant radial speed. Behaviours of outer shocked gas shell outside
the contact discontinuity are described by type $\mathcal{Z}$ ISSV
shock solutions with quasi-smooth edges. Stellar core collapses
prior to supernova explosions lead to neutrino bursts during a short
period of time, which might momentarily stand against the inward
pressure force across the `void' edge and give rise to a sharp-edge
`void' structure. In the long-term evolution after the escape of
neutrinos, diffusion effect and outer forward shocks will dominate
the behaviours of supernova remnants and the sharp edge will be
smoothed out eventually. In other situations when central magnetized
relativistic pulsar winds begin to resist diffusion effects, a
quasi-smooth void with shocked shell (i.e. type $\mathcal{Z}$ ISSV)
might also form.


Similar to PNe, interstellar bubbles may originate from strong
stellar winds of early-type stars on larger scales.
We
invoke type $\mathcal{Z}$ ISSV solutions with quasi-smooth void edge
to describe the structure and evolution of dense shell and outer ISM
envelope. In our model, the hot shocked stellar wind zone, which is
located between the reverse shock and the contact discontinuity
surface, is filled with steady shocked wind plasma. In Weaver et al.
(1977), the standard Spitzer conduction was included to study the
shocked stellar wind and shell gas that diffuse into the shocked
stellar wind region. However, the stellar magnetic field is
predominantly transverse to the radial direction at large radii,
which will suppress the thermal conduction through the hot
interstellar bubble gas (e.g., Chevalier \& Imamura 1983). As a weak
magnetic field can drastically reduce this thermal conduction
coefficient (e.g. Narayan \& Medvedev 2001; Malyshkin 2001) and a
weak magnetic field has little effects on behaviours of the gas
shells and nebulae (e.g. Avedisova 1972; Falle 1975), we do not
include thermal conduction effect in our model but present a dynamic
evolution model for interstellar bubbles in terms of a self-similar
nebular shell sustained by a central steady stellar wind. We would
note that the existence of a random magnetic field may reduce the
density 'wall' around a void and make the formation of a void easier
(Lou \& Hu 2009). We do not include magnetic field in our model for
simplicity.

%



Recent observations of ultraluminous X-ray sources (ULXs) show
that ULXs may blow very strong winds or jets into the surrounding
ISM and generate hot bubbles. For instance, ULX Bubble MH9-11
around Holmberg IX X-1 has experienced an average inflating
wind/jet power of $\sim 3\times 10^{39}$ erg s$^{-1}$ over an age
of $\sim 10^6$ years.
The shock of the bubble travels outwards at $\sim 100$ km s$^{-1}$
at radius $\sim 100$ pc. The particle density around the shock is
$\sim 0.3$ cm$^{-3}$ (e.g. Pakull \& Gris\'e 2008). Approximately,
this bubble corresponds to a type $\mathcal{Z}$ ISSV solution with
$\alpha_0\approx3.8\times 10^{-4}$, $x_0=0.4$ and $x_s=1$, and the
temperature of gas shell is $\sim 10^7$ K. Our ISSV model predicts
a contact discontinuity surface, or interaction surface of ULX
wind with the ISM, at a radius $\sim 40$ pc.


Finally, to diagnose voids observationally, we would suggest among
others to detect X$-$ray emissions from hot gas. In a companion
paper), we shall adapt our ISSV solutions for hot optically thin
X$-$ray gas clouds or nebulae, where we advance a useful concept
of projected self-similar X$-$ray brightness. It is possible to
detect ISSVs and identify diagnostic features of ISSV types which
may in turn to reveal clues of ISSV generation mechanisms. We will
also compare our ISSV model with observational data on more
specific terms. Moreover, projected self-similar X$-$ray
brightness is a general concept which can be useful when we
explore other self-similar hydrodynamic or magnetohydrodynamic
processes (e.g. Yu et al. 2006; Wang \& Lou 2008).

\begin{acknowledgements}
This research was supported in part by the National Natural Science
Foundation of China (NSFC) grants 10373009 and 10533020 at Tsinghua
University, the SRFDP 20050003088 and 200800030071, the Yangtze
Endowment and the National Undergraduate Innovation Training Project
from the Ministry of Education at Tsinghua University and Tsinghua
Centre for Astrophysics (THCA).
The kind hospitality of Institut f\"ur Theoretische Physik und
Astrophysik der Christian-Albrechts-Universit\"at Kiel Germany and
of International Center for Relativistic Astrophysics Network
(ICRANet) Pescara, Italy is gratefully acknowledged.
%
\end{acknowledgements}

\appendix

\section[]{Jump of $\alpha$ from zero to\\
\qquad\ \ nonzero value across the ZML}

Let us assume that the reduced mass density $\alpha(x)$ can transit
from zero to nonzero across the ZML and $x_0$ is the transition
point. For this, we require $v(x_0)=x_0$, $\alpha(x_0)=0$ and when
$x<x_0$, $\alpha(x)$ and $v(x)$ vanish. For an arbitrarily small
real number $\varepsilon>0$, we also require
$\alpha(x_0+\varepsilon)>0$ for a positive mass such that there
exists a positive integer $n$ for which
\begin{equation}
 \frac{d^n\alpha}{dx^n}\bigg|_{x_0}\neq0\ .
\end{equation}
We now cast equation (\ref{ODE2}) in the form of
\begin{equation}
\frac{d\alpha}{dx}=\alpha\frac{[\alpha-2(x-v)/x](x-v)}{(x-v)^2-1}\equiv
\alpha {\cal F}(x)\ ,
\end{equation}
where ${\cal F}(x)\equiv[\alpha-2(x-v)/x](x-v)/[(x-v)^2-1]$.

At $x=x_0$, the denominator of ${\cal F}(x)$ does not vanish because
$v(x_0)=x_0$. So ${\cal F}(x)$ is a finite continuous analytic
function near $x_0$. An arbitrary order derivative of ${\cal F}(x)$
at $x_0$ should be finite as well.

The $k$th-order derivative of equation (A2) by the Leibnitz rule
reads
\begin{equation}
  \frac{d^{k+1}\alpha}{dx^{k+1}}=\sum_{i=0}^kC_k^i\frac{d^{i}
  \alpha}{dx^i}\frac{d^{k-i}{\cal F}(x)}{dx^{k-i}}\ ,
\end{equation}
where $C_k^i$ stands for $k!/[i!(k-i)!]$ and $!$ is the standard
factorial operation.

Because $\alpha(x_0)=0$, equation (A3) yields $\alpha'(x_0)=0$,
$\alpha''(x_0)=0$ and so forth. That is, $\alpha(x_0)=0$ and
equation (\ref{ODE2}) determine that the arbitrary order derivative
of $\alpha(x)$ at $x_0$ is zero, which is contrary to presumption
(A1) based on the former assumption that $\alpha(x)$ can transit
from zero to nonzero across $x_0$ of the ZML. Therefore $\alpha$
cannot transit from zero to nonzero across the ZML for an isothermal
gas. For properties of void boundary in a polytropic gas, the
interested reader is referred to Hu \& Lou (2008) and Lou \& Hu
(2008).

\section[]{ Proof of inequality
$\alpha''(x_0)<0$}

Equation (\ref{ODE2}) can be written in the form of
\begin{equation}
\big[(x-v)^2-1\big]\alpha'=\alpha\big[\alpha-2(1-v/x)\big](x-v)\ ,
\end{equation}
whose first-order derivative $d/dx$ is simply
\begin{multline}
\!\!\!\!\!\!\!\!\!
2(x-v)(1-v')\alpha'+\big[(x-v)^2-1\big]\alpha''\\
\!\!\!\!\!\!\!\!\! =\{\alpha'\big[\alpha-2(1-v/x)\big]
+\alpha\big[\alpha'+2v'/x-2v/x^2\big]\}(x-v)\\
+\alpha\big[\alpha-2(1-v/x)\big](1-v')\ .
\end{multline}
We have $v(x_0)=x_0$ and thus both $v'(x_0)=0$ and $\alpha'(x_0)=0$
at $x_0$ by equation (\ref{voidedge}). Therefore at $x_0^{+}$,
equation (B2) becomes
\begin{equation}
\alpha''(x_0)=-\alpha(x_0)^2<0\
\end{equation}
for a positive density $\alpha(x_0)>0$ as a physical requirement.

\end{document}